\documentclass[aps,reprint,longbibliography]{revtex4-1}
\usepackage[utf8x]{inputenc}
\usepackage{amsmath,amssymb,amsfonts}
\usepackage{graphicx}
\usepackage{subfigure}
\usepackage{float}
\usepackage{color}

\newcommand{\wpl}{w^{+}}
\newcommand{\wmi}{w^{-}}

\newcommand{\op}{\Omega^{+}}
\newcommand{\om}{\Omega^{-}}
\newcommand{\fex}{f_{\text{ex}}}
\newcommand{\mean}[1]{\langle #1 \rangle}
\newcommand{\weq}{w_{0}}
\newcommand{\tp}{\theta^{+}}
\newcommand{\tm}{\theta^{-}}
\newcommand{\stot}{\dot{S}_{\text{tot}}}

\newcommand{\av}[1]{\left \langle #1 \right \rangle}
\newcommand{\tpa}{\theta^{\alpha,+}}
\newcommand{\tma}{\theta^{\alpha,-}}

\newcommand{\DF}{\Delta F^{\alpha}}
\newcommand{\T}{{\text{T}}}
\newcommand{\D}{{\text{D}}}
\newcommand{\Ph}{{\text{P}}}
\newcommand{\jix}{j_i^x(y)}

\begin{document}

\title{Effective rates from thermodynamically consistent coarse-graining of models for molecular motors with probe particles}

\author{Eva Zimmermann}%
\author{Udo Seifert}
   \affiliation{II. Institut f\"ur Theoretische Physik, Universit\"at Stuttgart, 70550 Stuttgart, Germany}%

\date{\today}

\begin{abstract}
Many single molecule experiments for molecular motors comprise not only the motor but also large probe particles coupled to it. The theoretical analysis of these assays, however, often takes into account only the degrees of freedom representing the motor. We present a coarse-graining method that maps a model comprising two coupled degrees of freedom which represent motor and probe particle to such an effective one-particle model by eliminating the dynamics of the probe particle in a thermodynamically and dynamically consistent way. The coarse-grained rates obey a local detailed balance condition and reproduce the net currents. Moreover, the average entropy production as well as the thermodynamic efficiency is invariant under this coarse-graining procedure. Our analysis reveals that only by assuming unrealistically fast probe particles, the coarse-grained transition rates coincide with the transition rates of the traditionally used one-particle motor models. Additionally, we find that for multicyclic motors the stall force can depend on the probe size. We apply this coarse-graining method to specific case studies of the $\text{F}_{1}$-ATPase and the kinesin motor.
\end{abstract}

\maketitle

\section{Introduction}

In many single molecule experiments beads that are attached to molecular motors are used to infer properties of the motor protein from the analysis of the trajectory of these probe particles. In particular, external forces can be exerted on the motor via such a probe particle \cite{kolo13,chow13}.
In the theoretical analysis of such assays, the motor is usually modelled as a particle hopping on a discrete state space with transitions governed by a master equation \cite{kolo07,lau07a,liep07a,liep08,lipo09a,astu10}. Alternatively, the so called ratchet models combine continuous diffusive spatial motion with stochastic switching between different potentials corresponding to different chemical states \cite{juel97,reim02a}.
These approaches often comprise only one particle explicitly, representing the motor.
The contribution of external forces which in the experiments act on the motor only via the probe are then included in the transition rates \cite{fish99,qian00a,kim05,liep07,liep07a,liep08,seif11a,golu12a,golu13} (or Langevin equation for the spatial coordinate \cite{gasp07,golu12}) of the motor particle directly. 
However, theoretical models that are used to reproduce the experimental observations should comprise at least two (coupled) degrees of freedom, one for the motor and one for the probe particle. Such models consisting of one degree of freedom hopping on a discrete state space representing the motor coupled to a continuously moving degree of freedom representing the probe are discussed in \cite{chen00,devi08,kunw08,korn09,bouz10,lade11,zimm12,piet14}. 
While multi-particle models are more precise and better represent the actual experimental setup, one-particle models are widely-used toy models often applied to illustrate basic ideas. 

Simplifying the description of systems consisting of many degrees of freedom with a concomitant large state space while still maintaining important properties is commonly known as coarse-graining. In the context of stochastic thermodynamics \cite{seif12}, various coarse-graining methods have been applied, e.g., lumping together states of a discrete state space among which transitions are fast \cite{raha07,nico11c,espo12,bo14}, averaging over states for discrete \cite{sant11} or continuous processes \cite{kawa12,bo14}, eliminating single states from a network description \cite{pigo08,pugl10,alta12} or eliminating slow (invisible) degrees of freedom \cite{aman10,mehl12,cris12}. It was found that, in general, coarse-graining has implications on the entropy production and, in particular \cite{gome08a}, dissipation. In the context of biological systems and especially molecular motors, coarse-graining procedures mostly focus on eliminating selected states of the motor \cite{tsyg08,alta12} or on reducing continuous (ratchet) models to discrete-state models \cite{kell00,latt02,latt04,gerr10,kram10}.

In the present paper, we introduce a coarse-graining procedure that allows to reduce molecular motor-bead models to effective one-particle models with discrete motor states with the external force acting directly on the effective motor particle. We eliminate the explicit dynamics of the probe particle completely still maintaining the correct local detailed balance condition for the effective motor transition rates and preserving the average currents of the system. As a main result, we find that the coarse-grained rates show a more complex force dependence than the usually assumed exponential behaviour and a more complex concentration dependence than mass action law kinetics. 

The paper is organized as follows. In section \ref{sec:onestate}, we introduce our coarse-graining method on the basis of a simple motor-bead model with only one motor state and apply it to a model for the $\text{F}_{1}$-ATPase \cite{zimm12}. In section \ref{sec:multistate}, we generalize the procedure to motor models with several internal states and apply it to both a refined model for the $\text{F}_{1}$-ATPase and to a kinesin model. A possible experimental implementation of our method is presented in section \ref{sec:exp}. We show that entropy production and efficiency remain invariant under this coarse-graining procedure in section \ref{sec:entropy}, discuss implications on the stall conditions in section \ref{sec:stall} and conclude in section \ref{sec:conc}.

\section{General one-state motor model}\label{sec:onestate}
\subsection{Explicit motor-bead dynamics}

\begin{figure}[ht]
\centering
\includegraphics[width=1.0\linewidth]{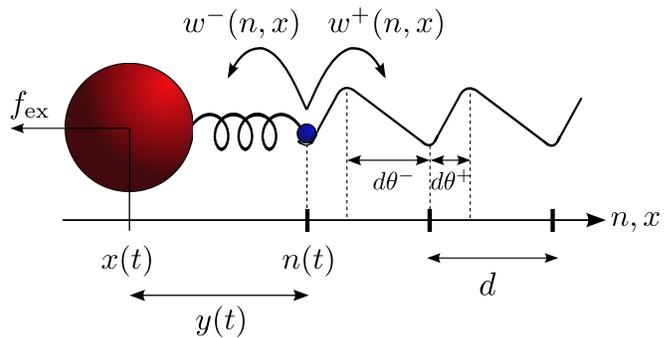}
\caption{(Color online) Schematic representation of a motor-bead model comprising a one-state motor (small blue sphere) attached via an elastic linker to the probe particle (large red sphere). An external force $\fex$ is applied to the bead. The transition rates of the motor are denoted by $\wpl(n,x)$ and $\wmi(n,x)$. The load sharing factors $\tp$ and $\tm$ indicate the position of an underlying unresolved potential barrier relative to the minimum of the free energy landscape of the motor.}
\label{fig:model}
\end{figure}

The general model for motor proteins with only one chemical state consists of one degree of freedom representing the motor which jumps between discrete states $n(t)$ separated by a distance $d$. 
The motor is coupled with the second degree of freedom representing the probe particle via some kind of elastic linker, see Fig. \ref{fig:model} \cite{zimm12}. The motion of the probe particle with continuous coordinate $x(t)$ is described by an overdamped Langevin equation with friction coefficient $\gamma$ and constant external force $\fex$,
\begin{align}
 \dot{x}(t)=\left(-\partial_x V(n-x)-\fex\right)/\gamma+\zeta(t),
\end{align}including the potential energy of the linker $V(n-x)$ and thermal noise $\zeta(t)$ with correlations $\langle\zeta(t_2)\zeta(t_1)\rangle=2\delta(t_2-t_1)/\gamma$. Throughout the paper, we set $k_{\text{B}}T=1$. This choice implies that the product of force $\fex$ and distance $d$ appearing in the figures below is measured in units of $k_{\text{B}}T$. The (instantaneous) distance between motor and probe is denoted by $y$. The system is characterized by the pair of variables ($n$,$x$) and is ``bipartite'' in these variables since transitions do not happen in both variables at the same time. The transition rates of the motor fulfill a local detailed balance (LDB) condition 
\begin{align}
 \frac{\wpl(y)}{\wmi(y+d)}=\exp[\Delta\mu-V(y+d)+V(y)].
\end{align}
The free energy change of the solvent $\Delta\mu\equiv\mu_\T-\mu_\D-\mu_\Ph$ with $\mu_i=\mu_i^{\text{eq}}+\ln(c_i/c_i^{\text{eq}}) $ and nucleotide concentrations $c_i$ is associated with ATP turnover.
The probability density $p(y)$ for the distance $y$ obeys a Fokker-Planck-type equation
\begin{align}
\partial_t p(y)& = \partial_y\left( \left(\partial_y\,V(y)-\fex\right)\,p(y)+\partial_y\,p(y)\right)/\gamma  \nonumber \\
	  &+ \wpl(y-d)\,p(y-d)+\wmi(y+d)\,p(y+d) \nonumber \\
	  &-\left(\wpl(y)+\wmi(y)\right)\,p(y).
\label{FPE}
\end{align}
For constant nucleotide concentrations, the system reaches a non-equilibrium stationary state (NESS) with constant average velocity 
\begin{align}
v&\equiv d \int_{-\infty}^{\infty} p^s(y)(\wpl(y)-\wmi(y)) \, \mathrm{d}y \label{eq:vel}\\ &=\int_{-\infty}^{\infty} p^s(y) \left(\partial_y V(y)-\fex\right)/\gamma \, \mathrm{d}y \nonumber
\end{align}
and stationary distribution $p^s(y)$.

\subsection{Coarse-graining procedure}
\label{sec:CG-one}

In the coarse-grained description of the model we want to map the motor-bead system to one effective motor particle hopping between states separated by $d$. We thus have to eliminate the $x$-coordinate from the ($n$,$x$)-description resulting in a system characterized only by $n$. 

For the coarse-grained model, we impose the following conditions. The coarse-grained transition rates $\Omega^{\pm}$ which advance the effective particle by $d$ should obey a LDB condition
\begin{align}
 \frac{\op}{\om}=\exp[\Delta\mu-\fex d]
\label{ldbcond}
\end{align}
as the force is now assumed to act directly on the effective motor particle. Furthermore, we require that the coarse-grained particle moves with the same average velocity in the steady state as the motor and the probe in the original model, i.e., 
\begin{align}
v=d(\op-\om).
\label{vcond}
\end{align} 
Solving the linear system of equations (\ref{ldbcond}, \ref{vcond}) yields the coarse-grained rates
\begin{align}
 \op&=\frac{v\exp[\Delta\mu-\fex d]/d}{\exp[\Delta\mu-\fex d]-1} \label{OP} \\
 \om&=\frac{v/d}{\exp[\Delta\mu-\fex d]-1}. \label{OM} 
\end{align}

The coarse-grained rates can be interpreted as effective transition rates that correspond to a transition process after which both particles, motor and probe, have advanced a distance $\pm d$. In principle, there are (for any $y$) many possible displacement processes to advance both particles by $d$, including ones with $l$ forward and $l-1$ backward motor jumps. The coarse-grained rate corresponds to the rate with which one such effective displacement will happen.

In general, the coarse-grained rates depend (via $v$) on all model parameters, including the friction coefficient of the probe particle and the specific potential of the linker.
If one had chosen coarse-grained rates by just averaging over the positions of the probe particle, i.e., by
\begin{align}
 \langle w^{\pm}\rangle=\int_{-\infty}^{\infty}  p^s(y) w^{\pm}(y)\, \mathrm{d}y, \label{Oav}
\end{align}
one would have obtained rates that yield the correct average velocity but do not fulfill the LDB condition, as discussed in section \ref{sec:cg-av} below.

For a more explicit analysis, we must specify the forward and backward rates of the motor. We choose \cite{zimm12}
\begin{align}
\wpl(y)&=w_0\exp[\mu^{+}-V(y+d\tp)+V(y)] \label{wplus}\\
\wmi(y)&=w_0\exp[\mu^{-}-V(y-d\tm)+V(y)] \label{wminus}
\end{align}
where $\tp$ and $\tm$ are the load-sharing factors with $\tp+\tm=1$ and $\mu^{+}=\mu_\T$, $\mu^{-}=\mu_\D+\mu_\Ph$. We assume an exponential dependence of the transition rates on the potential difference of the linker according to Kramers' theory. This exponential dependence on the potential difference is similar to one-particle models where the rates of the motor typically depend exponentially on the external force with a corresponding load-sharing factor \cite{kolo07,liep07a}.

\subsection{Time-scale separation}

In this section, we will investigate under which conditions the coarse-grained rates (\ref{OP}, \ref{OM}) can be expressed using a single exponential dependence on the external force as typically assumed for mechanical transitions within one-particle models \cite{liep07a,kolo07}. 

Inserting eqs. (\ref{wplus}, \ref{wminus}) in eq. (\ref{FPE}) in the NESS shows that the contribution due to motor jumps is weighted with a (dimensionless) prefactor 
\begin{align}
\varepsilon\equiv\weq\exp[\mu_\T^{\text{eq}}] d^2\gamma.
\end{align}
Here, $\weq\exp[\mu_\T^{\text{eq}}]$ determines the timescale of the transitions of the motor while $\gamma d^2$ determines the timescale of the dynamics of the probe particle. The latter is mainly governed by the size of the bead and the step size of the motor whereas $\weq\exp[\mu_\T^{\text{eq}}]$ is determined by the attempt frequency and also by the absolute nucleotide concentrations.

If the dynamics of the bead is much faster than the transitions of the motor, time-scale separation holds with $\varepsilon\rightarrow 0$ \cite{qian00b,espo12}. In this limit of fast bead relaxation, denoted throughout by a caret, the stationary solution of eq. (\ref{FPE}) in the NESS becomes
\begin{align}
 \hat{p}^{s}(y)=\exp[-V(y)+\fex y]/\mathcal{N} 
\end{align}
with $\mathcal{N}\equiv\int_{-\infty}^{\infty}\exp[-V(y)+\fex y]\,\mathrm{d}y$. The average velocity is then given by
\begin{align}
 \hat{v}=&d\int_{-\infty}^{\infty} \hat{p}^{s}(y)(\wpl(y)-\wmi(y)) \, \mathrm{d}y \nonumber \\
  =&d\weq\left(e^{\mu_\T-\fex d\tp}- e^{\mu_\D+\mu_\Ph+\fex d\tm}\right).
\end{align}
This expression inserted into eqs. (\ref{OP}, \ref{OM}) yields
\begin{align}
 \hat{\Omega}^{+}&=\weq e^{\mu_\T-\fex d\tp}, \label{OPmu}\\
 \hat{\Omega}^{-}&=\weq e^{\mu_\D+\mu_\Ph+\fex d\tm} \label{OMmu}
\end{align}
independent of any specific linker potential $V(y)$. Since this force dependence is purely exponential with the correct load sharing factor, these expressions represent exactly the rates typically used in one-particle models. We notice that within this approximation $\op=\mean{\wpl(y)}$ and $\om=\mean{\wmi(y)}$ holds true, which is in agreement with other coarse-graining procedures in the time-scale separation limit, e.g., \cite{sant11,espo12,bo14}.

Note that only transition rates of the motor whose dependence on the linker potential is chosen accordingly in the Kramers form (eqs. (\ref{wplus}, \ref{wminus})) lead generically to consistent coarse-grained and averaged rates when using the fast-bead limit of $p^s(y)$.

\subsection{Example: $\text{F}_{\boldsymbol 1}$-ATPase}
\label{sec:F1120}

In general, a strong time-scale separation between motor and probe is not necessarily realistic. In this case, eq. (\ref{FPE}) must be solved numerically. We will use the model introduced in \cite{zimm12}, see Fig. \ref{fig:model}, with a harmonic potential $V(y)=\kappa y^2/2$ as a simple example to illustrate our coarse-graining procedure.

\begin{figure}[ht]
\centering \subfigure{\includegraphics[width=1.0\linewidth]{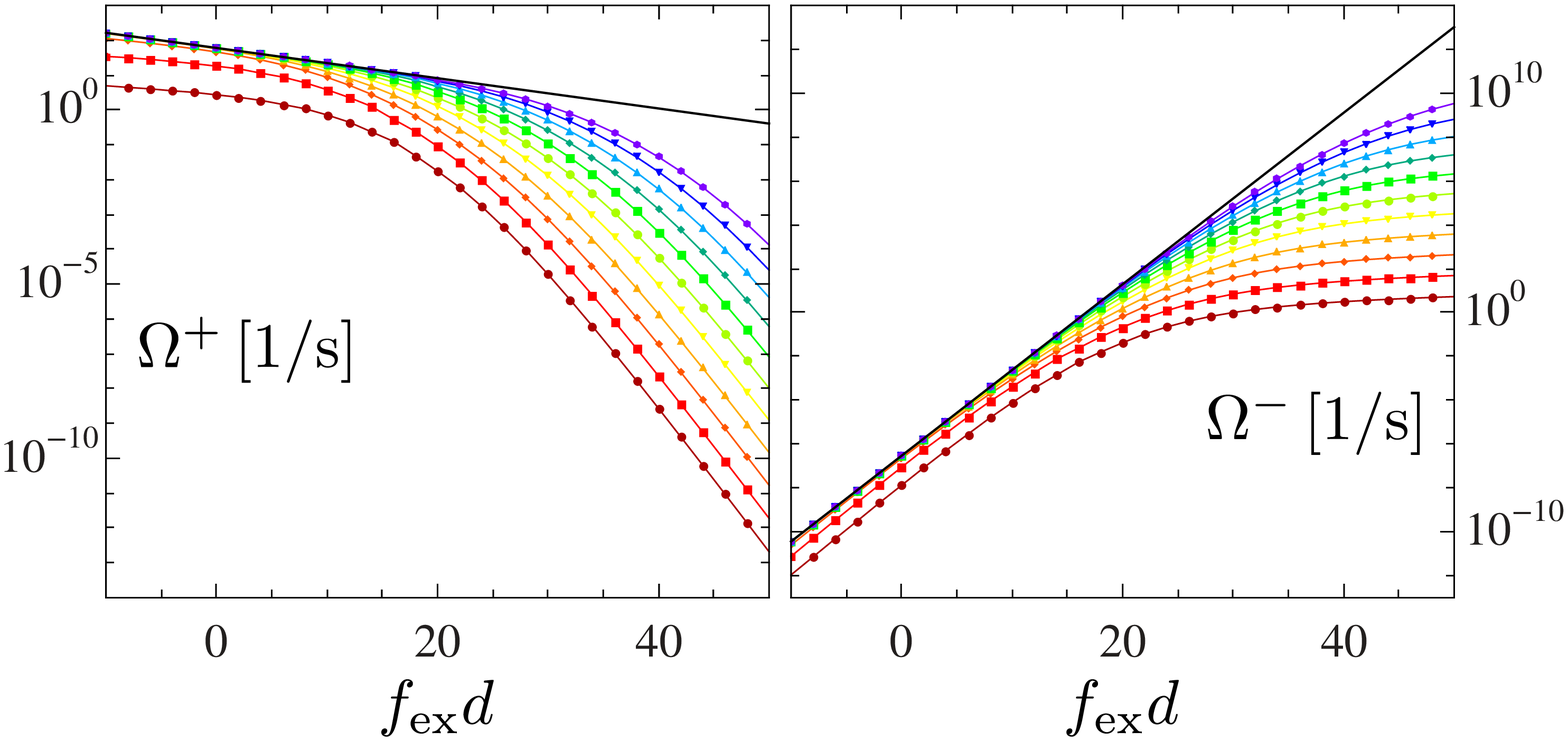}} 
\raggedright \subfigure{\includegraphics[width=0.95\linewidth]{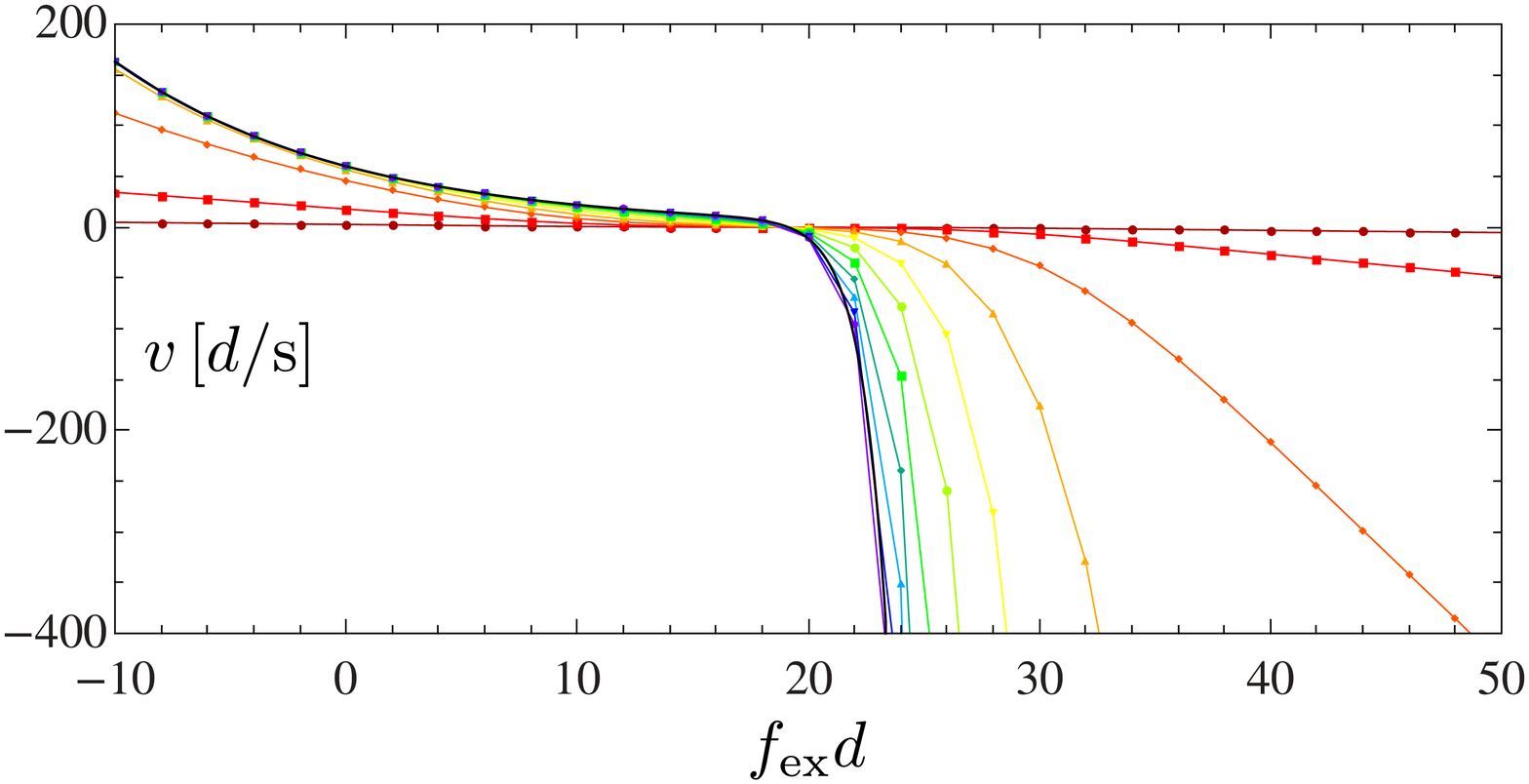}}
\caption{(Color online) Coarse-grained rates $\op$ and $\om$ (top) and average velocity (bottom) as functions of $\fex d$ for various friction coefficients $\gamma$ in the range $5\,\text{s}/d^2 \, \geq \gamma \, \geq \,\,5\cdot 10^{-10}\,\text{s}/d^2$ (from bottom to top). With decreasing $\gamma$, the rates and the velocity approach the corresponding fast-bead limit (solid black lines). Parameters: $\kappa=40 d^{-2}$, $c_\T=c_\D=2\cdot10^{-6}$M, $c_\Ph=10^{-3}$M, $\Delta\mu=19$, $\tp=0.1$, $w_0\exp[\mu_\T^{\text{eq}}]/c_\T^{\text{eq}}=3\cdot 10^{7}(\text{Ms})^{-1}$.}
\label{OPOM}
\end{figure}

In Fig. \ref{OPOM}, the results for $\op$ and $\om$ are shown for various values of the friction coefficient $\gamma$. With decreasing $\gamma$, the rates approach their corresponding fast-bead limits, $\hat{\Omega}^{+}$ and $\hat{\Omega}^{-}$. These values are upper bounds because decreasing $\gamma$ implies smaller probe particles which exert less drag on the motor. For finite $\gamma$, the coarse-grained rates do not show a single exponential dependence on $\fex$ over the whole range of external forces. Such a dependence, however, is usually assumed to hold within one-particle models. Moreover, the coarse-grained rates depend on $\gamma$, which is a parameter not incorporated explicitly in many one-particle models.

The experimentally accessible values of $\gamma$ cover a wide range of the values chosen in Fig. \ref{OPOM}. A dimer of polystyrene beads ($\simeq 280$ nm) as used in \cite{toya10,haya10,toya11,wata13} corresponds to $\gamma=0.5\,\text{s}/d^2$ (red (dark gray) line with squares) while a $40\,\text{nm}$-gold particle \cite{yasu01,adac07,wata13} corresponds to $\gamma=5\cdot10^{-4}\,\text{s}/d^2$ (yellow (light gray) line with triangles). Especially for large external forces, the coarse-grained rates deviate strongly from their asymptotic values even for a probe as small as the gold particle. 

The average velocity as shown in Fig. \ref{OPOM} also strongly depends on the friction coefficient of the probe particle, especially for large external forces. In this regime, for large $\gamma$, the velocity is dominated by the friction experienced by the probe while for small $\gamma$ the probe relaxes almost immediately and the velocity is dominated by the timescale of the motor jumps.

\begin{figure}[t]
  \centering
\includegraphics[width=1.0\linewidth]{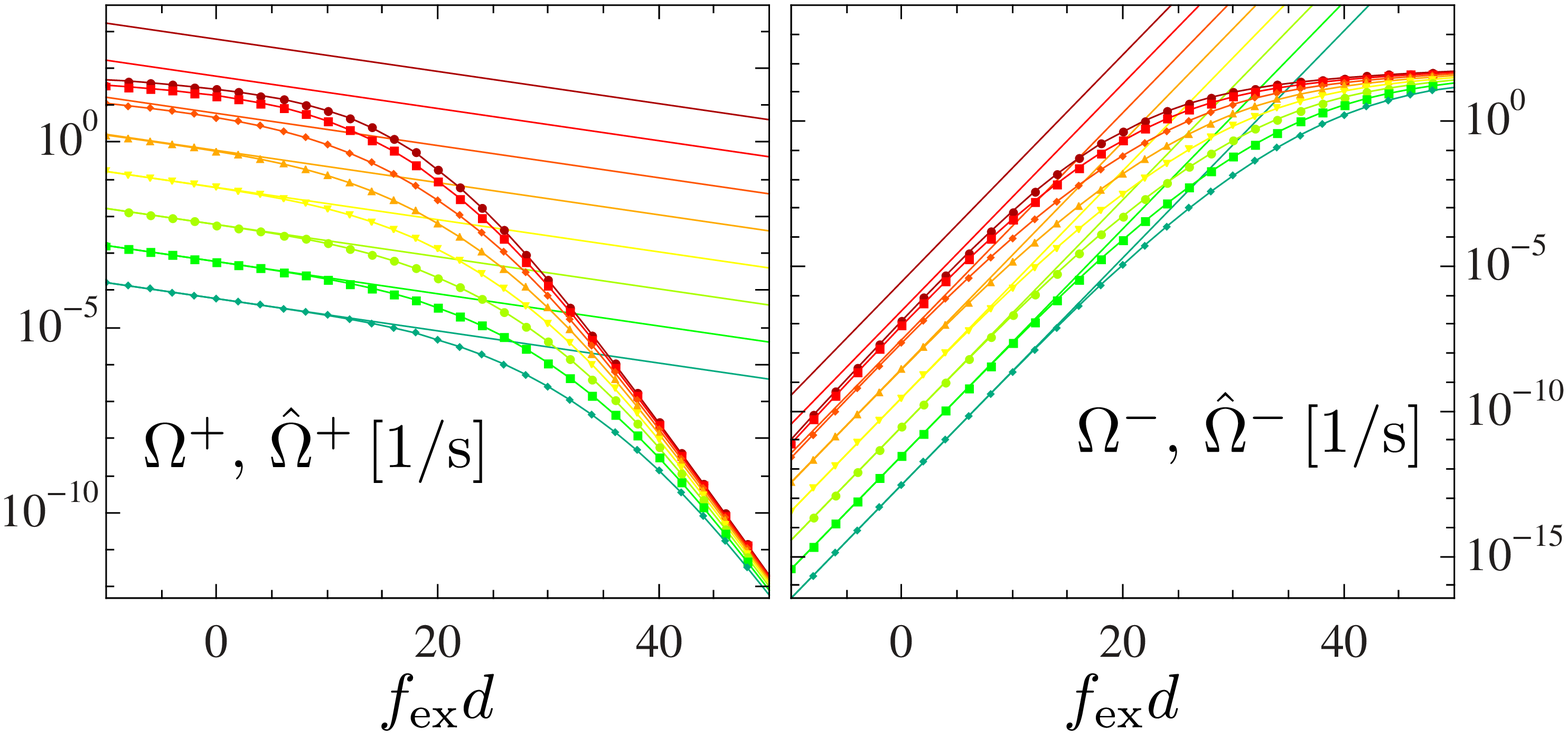}
\caption{(Color online) Coarse-grained rates $\op$ and $\om$ as functions of $\fex d$ for various $c_\T$, $c_\D$ in the range $2\cdot10^{-5}\,\text{M} \, \geq c_\T,c_\D \, \geq \,\,2\cdot 10^{-12}\,\text{M}$ (from top to bottom). With decreasing $c_\T,c_\D$, the rates approach the fast-bead limits $\hat{\Omega}^{+}$ and $\hat{\Omega}^{-}$ (straight lines). Parameters: $\kappa=40 d^{-2}$, $\gamma=0.5\text{s}/d^2$, $c_\Ph=10^{-3}$M, $\Delta\mu=19$, $\tp=0.1$, $w_0\exp[\mu_\T^{\text{eq}}]/c_\T^{\text{eq}}=3\cdot 10^{7}(\text{Ms})^{-1}$.}
\label{OPOMct}
\end{figure}

Another option to reach the fast-bead limit is to use very small nucleotide concentrations. In Fig. \ref{OPOMct}, we show the coarse-grained rates for various ATP and ADP concentrations. With decreasing nucleotide concentration (at fixed $\Delta \mu$), the rates approach the asymptotic $\hat{\Omega}^{+}$ and $\hat{\Omega}^{-}$. However, it is very hard to do experiments at concentrations smaller than $\simeq 10^{-7}$M as jumps of the motor are then very rare.

In Fig. \ref{OPOM} and in Fig. \ref{OPOMct} the dependence of the coarse-grained rates on the external force exhibits two different regimes. Up to values of the external force of roughly $15 /d$, the coarse-grained rates can be well approximated by a single exponential dependence on $\fex$ with the same slope as in the fast-bead limit, $d\tp$ or $d\tm$, respectively. However, for large $\gamma$ and large $c_\T$, even in this regime, the absolute values of the coarse-grained rates deviate up to two orders of magnitude from their fast-bead approximation. For such parameters, assuming a mono-exponential dependence on $\fex$ with the above slope would not be appropriate either.

For large external forces, all coarse-grained rates deviate significantly from their fast-bead limits. We find again a mono-exponential decay for $\op$ but now with slope $-d$ whereas $\om$ grows only linearly with increasing $\fex$. This so far unaccounted for behaviour can be understood by considering the limit $\fex\rightarrow\infty$ as discussed in detail in the Appendix. The crossover from one regime to the other occurs beyond the stall force $\fex=\Delta\mu/d$.

In summary, we find that for the $\text{F}_1$-ATPase under realistic experimental conditions the rates in a coarse-grained description comprising only one effective particle that satisfy the LDB condition eq. (\ref{ldbcond}) and reproduce the correct average velocity $v$ can not be written in the form of a single exponential dependence on the external force.

\subsection{Comparison of coarse-grained with averaged rates}
\label{sec:cg-av}

\begin{figure}[t]
\hfill \subfigure{\includegraphics[width=0.98\linewidth]{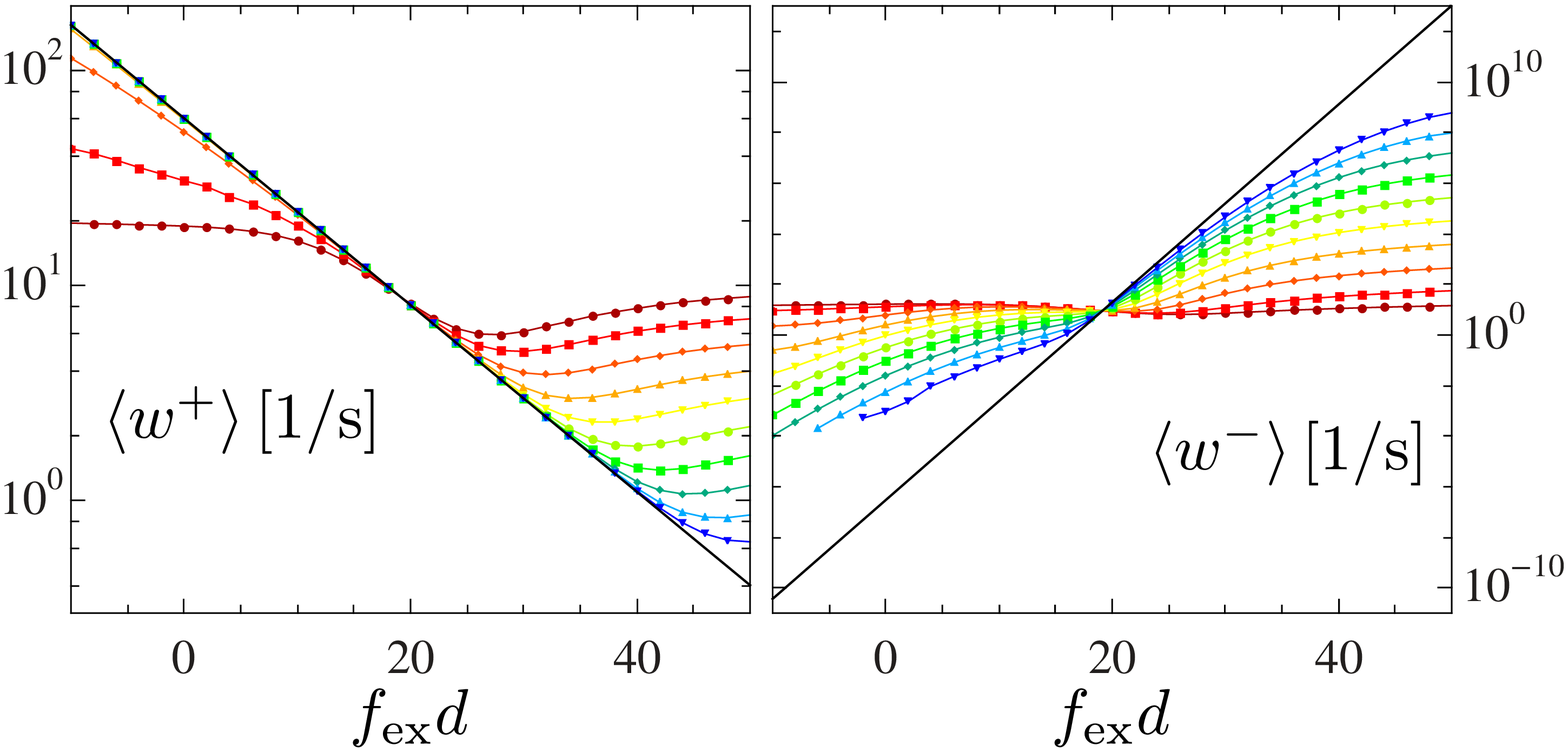}}
\raggedright \subfigure{\includegraphics[width=0.95\linewidth]{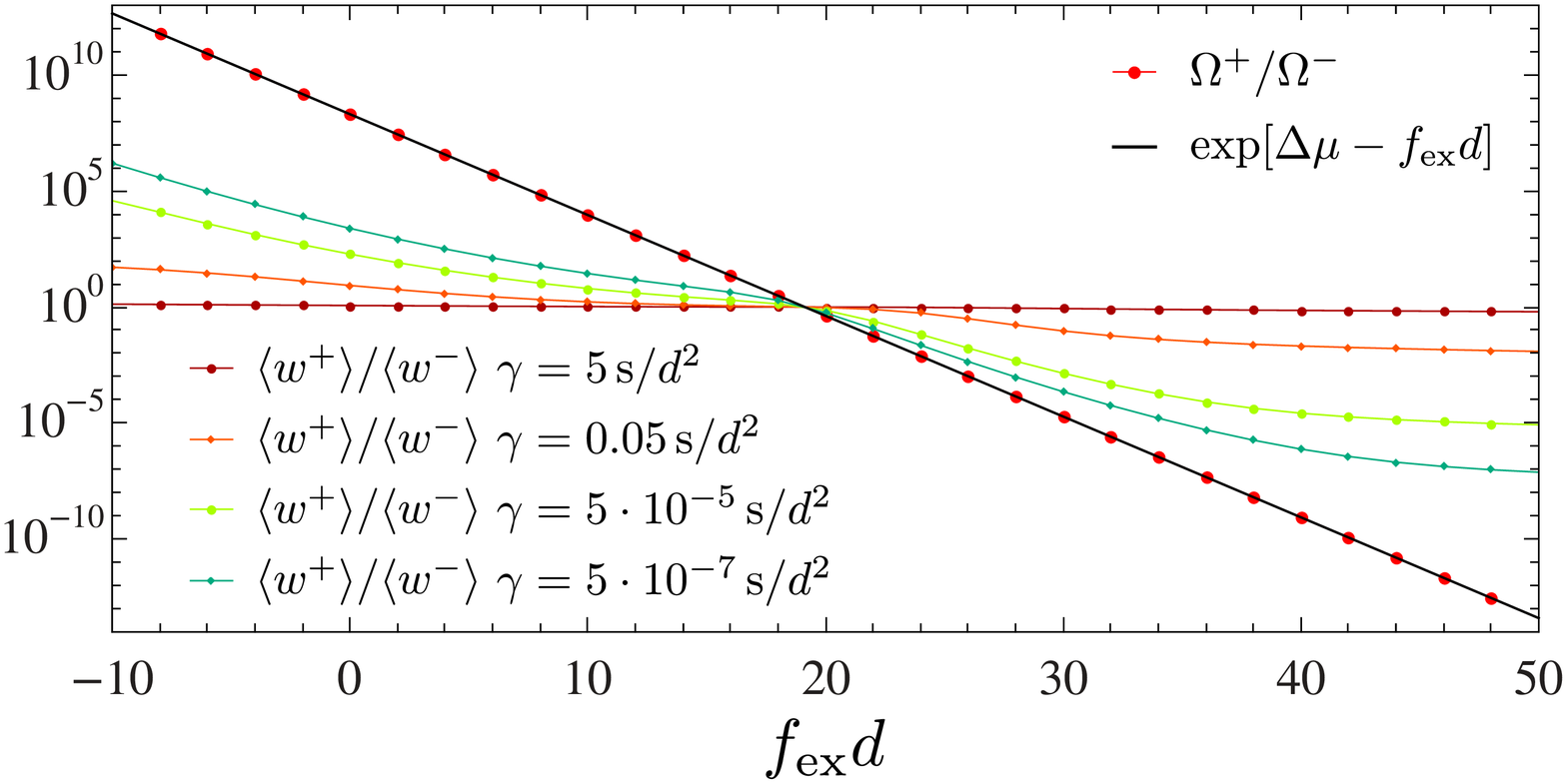}}
\caption{(Color online) Top: Average rates $\langle\wpl\rangle$ and $\langle\wmi\rangle$ as functions of $\fex d$ for various $\gamma$ in the range $5\,\text{s}/d^2 \, \geq \gamma \, \geq \,\,5\cdot 10^{-9}\,\text{s}/d^2$. With decreasing $\gamma$, the rates approach $\hat{\Omega}^{+}$, $\hat{\Omega}^{-}$ (solid black lines). Bottom: Ratio of $+$ and $-$ rates. In contrast to $\op$, $\om$ (large red dots), the averaged motor rates do not fulfill the LDB condition (solid black line). The parameters are the same as in Fig. \ref{OPOM}.}
\label{120F1ldb}
\end{figure}

Instead of defining the coarse-grained rates according to eqs. (\ref{OP}, \ref{OM}), one might be tempted to use the averaged rates (\ref{Oav}) as a definition for the coarse-grained rates. In Fig. \ref{120F1ldb}, we show the averaged rates of our $\text{F}_1$-ATPase model as well as their ratio corresponding to the LDB condition. We find that both $\langle\wpl\rangle$ and $\langle\wmi\rangle$ (for the latter less visible in the plot) exhibit non-monotonic dependence on the external force. For external forces slightly larger than the stall force, $\langle\wpl\rangle$ increases with increasing $\fex$ due to the fact that in this region the system moves backward with motor jumps following the probe which leads to a peak at small $y$ in $p^s(y)$. On the other hand, $\langle\wmi\rangle$ exhibits a minimum around stall conditions for large $\gamma$ since in this region, $p^s(y)$ misses a peak at large $y\simeq 1$.

A severe issue appears regarding the LDB condition. The corresponding ratio of the averaged rates is also plotted in Fig. \ref{120F1ldb} where it can be clearly seen that the LDB condition is not fulfilled (except in the fast-bead limit).

\subsection{Without external force}

Even though we have motivated this paper by emphasizing that external forces are typically applied to probe particles, it should be obvious that our approach holds true for molecular motors transporting cargo subject to Stokes friction in the absence of external forces.

For one-particle models, the friction coefficient of the probe can not be taken into account explicitly. One rather has to incorporate the drag effect of the bead into the motor rates \cite{gerr10}. If one wants to analyze experimental data obtained from probe particles of different sizes, one then has to use different values of the motor rates for each data set.

For the rather dilute solutions used in experiments \cite{cart05,toya10,toya11} one generally assumes that the motor dynamics is subject to mass action law kinetics, i.e., that the transition rates depend linearly on the corresponding concentration of nucleotides. Obviously, this linear dependence holds for all concentrations and beads of all sizes for one-particle models. When keeping $c_\D$ and $c_\Ph$ fixed, the average velocity of a one-state motor will show a purely linear dependence on $c_\T$.

The experimental analysis of the average velocity of the $\text{F}_{1}$-ATPase as function of $c_\T$ (for fixed $c_\D$, $c_\Ph$) reveals a saturation of the velocity for large ATP concentrations which sets in earlier for large beads \cite{yasu01}. While such a saturation is usually attributed to the hydrolysis step, we find that a sub-linear dependence of the velocity can also be caused by the drag of the probe particle.

\begin{figure}[t]
\centering \subfigure{\includegraphics[width=1.0\linewidth]{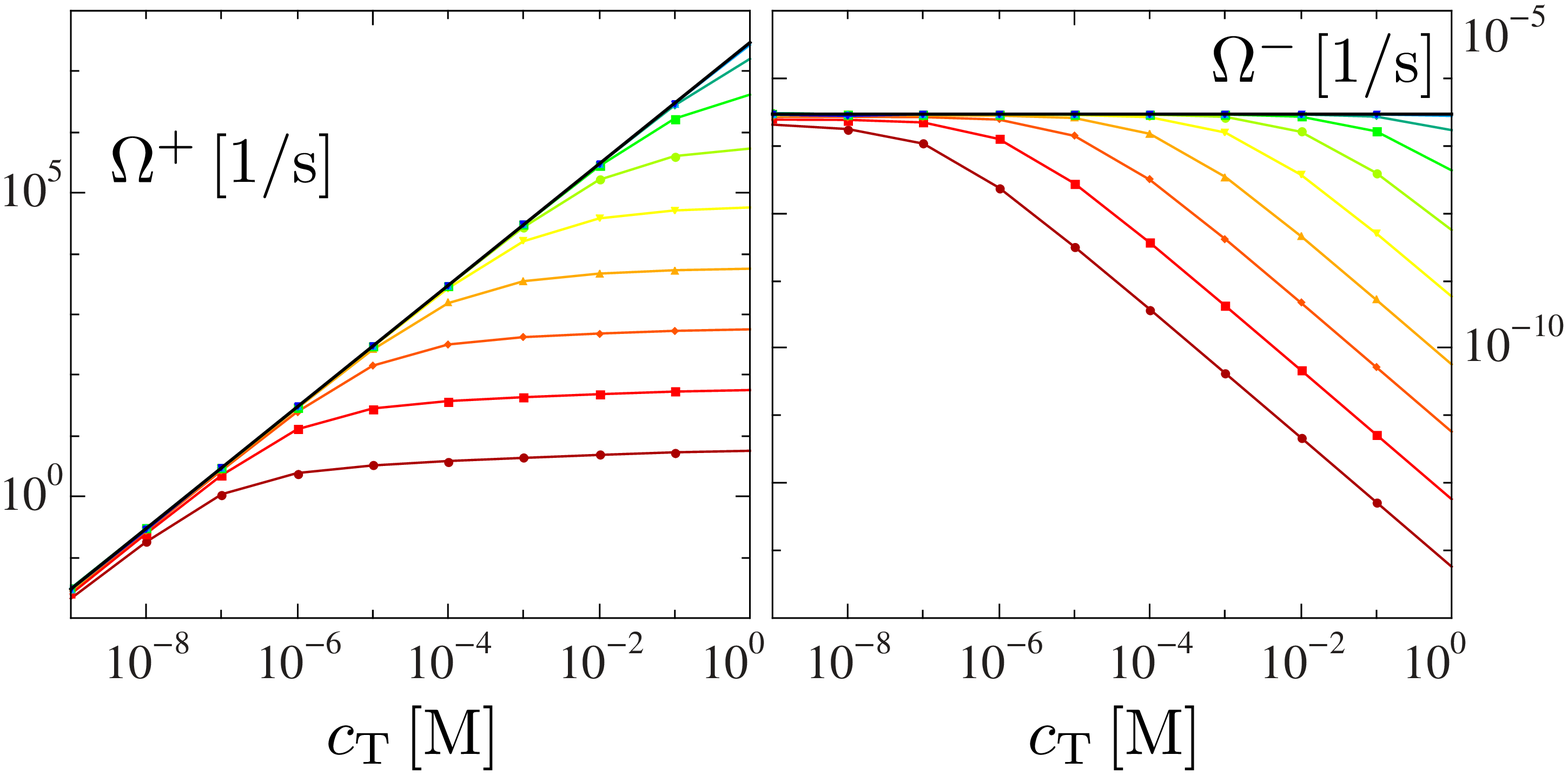}}
\raggedright \subfigure{\includegraphics[width=0.96\linewidth]{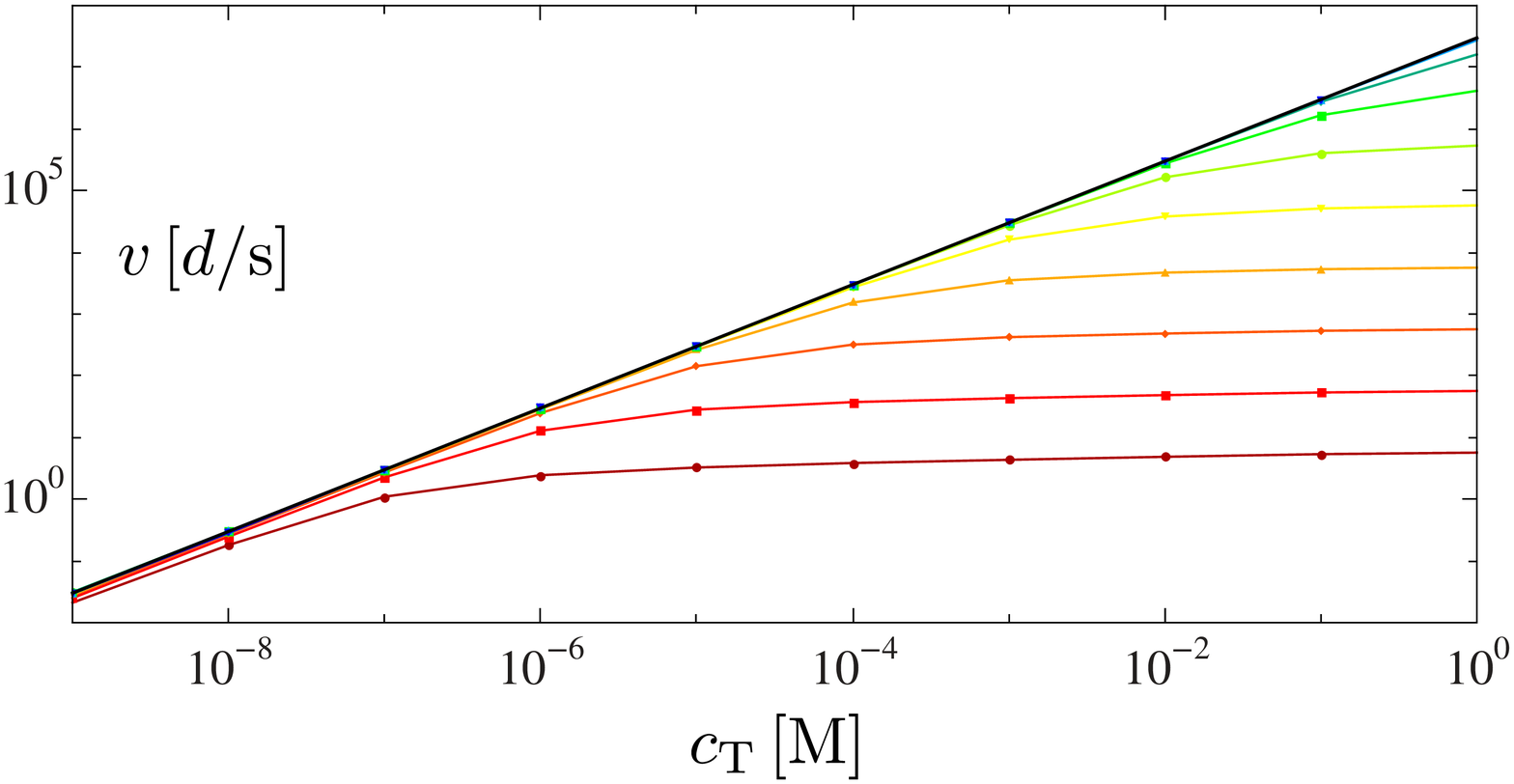}}
\caption{(Color online) Coarse-grained rates $\op$ and $\om$ (top) and average velocity (bottom) for various $\gamma$ and $\fex=0$ as functions of $c_\T$. Since $c_\D$ and $c_\Ph$ are fixed, $\Delta\mu$ also increases with $c_\T$. The rates and the velocity approach the fast-bead approximation (solid black lines). Parameters: $c_\D=2\cdot10^{-6}\text{M}$, $c_\Ph=1\cdot10^{-3}\text{M}$, $\kappa=40 d^{-2}$, $\tp=0.1$, $w_0\exp[\mu_\T^{\text{eq}}]/c_\T^{\text{eq}}=3\cdot 10^{7}(\text{Ms})^{-1}$, $\gamma$ in the range $5\,\text{s}/d^2 \, \geq \gamma \, \geq \,\,5\cdot 10^{-9}\,\text{s}/d^2$ (from bottom to top). }
\label{fig:gamma120}
\end{figure}

In Fig. \ref{fig:gamma120}, the coarse-grained rates as well as the velocity are shown as a function of the ATP concentration. With decreasing $\gamma$, the coarse-grained rates approach the fast-bead limit and the mass action law kinetics. The velocity is then linear in $c_\T$ as in a one-particle model. For large $\gamma$, eliminating the cargo by coarse-graining yields coarse-grained rates that are not linear in the concentrations although the motor rates are still subject to mass action law kinetics. Moreover, the velocity then exhibits a sub-linear dependence reminiscent of the typical saturation effect for large $c_\T$.

\subsection{Comparison of full and coarse-grained trajectories}
\label{ill}

\begin{figure}[t]
 \centering\includegraphics[width=1.0\linewidth]{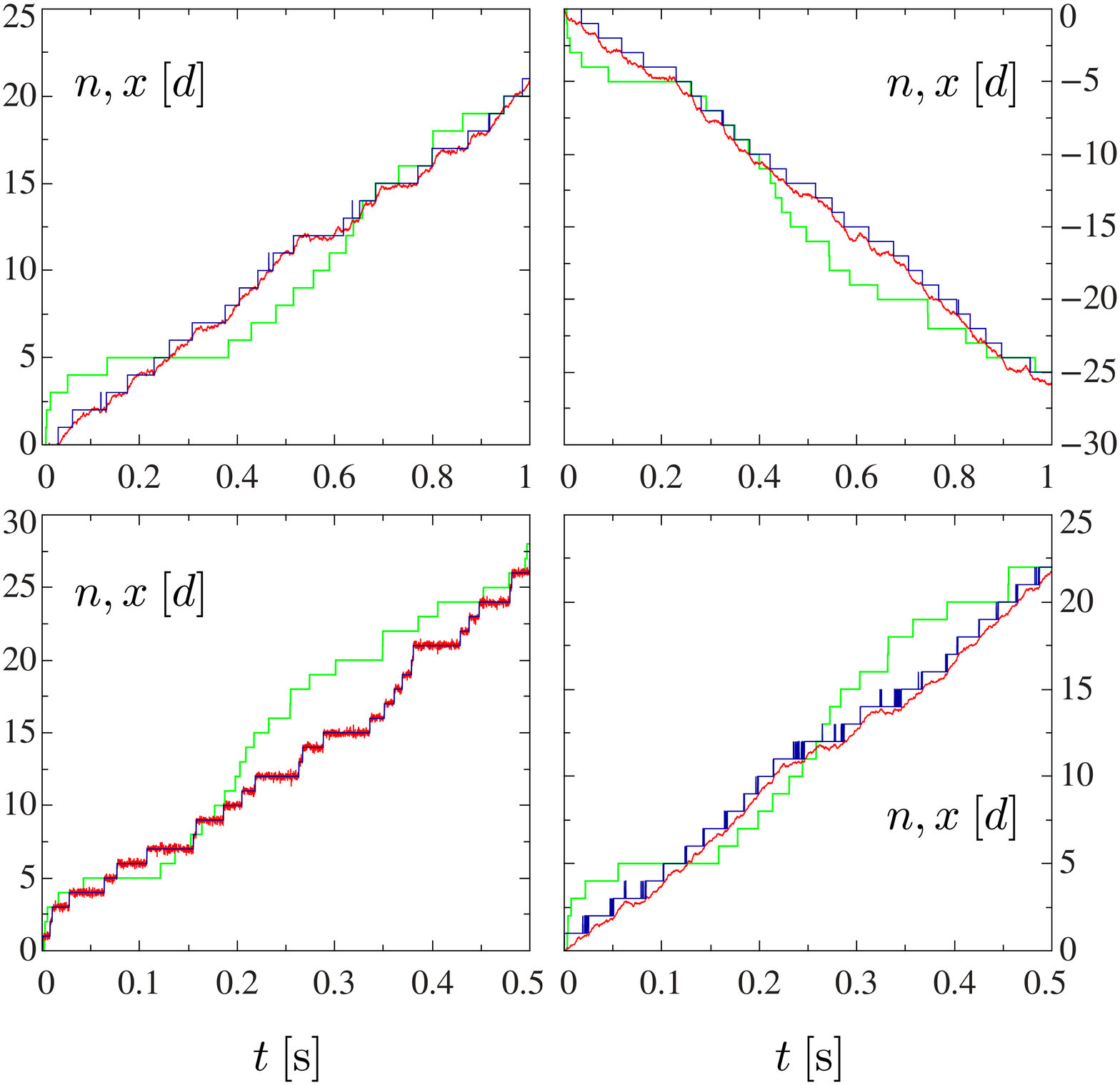}
\caption{(Color online) Trajectories of the one-state model for the $\text{F}_{1}$-ATPase for several parameter sets obtained from simulations. The trajectory of the detailed model (motor: step-like blue lines, probe: fluctuating red lines) is shown together with a trajectory of its corresponding coarse-grained model (green (light gray)). Parameters: $\kappa=40d^{-2}$, $\tp=0.1$, $\weq\exp[\mu_\T^{\text{eq}}]/c_\T^{\text{eq}}=3\cdot 10^7$ $(\text{Ms})^{-1}$, $\gamma=0.5\,\text{s}/d^2$, $\fex=0$, $c_\T=c_\D=2\cdot 10 ^{-6}$ M, $c_\Ph=0.001$ M (top left); $\gamma=0.5\,\text{s}/d^2$, $\fex=40\,d^{-1}$, $c_\T=c_\D=2\cdot 10 ^{-6}$ M, $c_\Ph=0.001$ M (top right); $\gamma=0.005\,\text{s}/d^2$, $\fex=0$, $c_\T=c_\D=2\cdot 10 ^{-6}$ M, $c_\Ph=0.001$ M (bottom left); $\gamma=0.5\,\text{s}/d^2$, $\fex=0$, $c_\T=0.001$ M, $c_\D=2\cdot 10 ^{-6}$ M, $c_\Ph=0.001$ M (bottom right).}
\label{traj}
\end{figure}

Trajectories of motor and probe generated by a simulation of the complete model of the $\text{F}_{1}$-ATPase are shown in Fig. \ref{traj}. Additionaly, Fig. \ref{traj} contains a trajectory obtained from simulating the corresponding coarse-grained model. The average velocity of both models is the same (by definition, see eq. (\ref{vcond})), whereas the coarse-grained model produces trajectories that are ``more random''. This behavior occurs since the coarse-grained rates are constant (for fixed parameters) and produce a simple biased random walk. The motor transition rates of the complete model, however, depend on the actual position of the probe and are therefore implicitly time-dependent. Since fast successive motor jumps are suppressed, the trajectory of the complete model is less random \cite{schi06,devi08}.

The influence of parameters like the probe size or the ATP concentration on the dynamics is visible in the bottom panels of Fig. \ref{traj}. While the average velocity is almost the same, the trajectories of the complete model differ significantly. Using a small probe with a small friction coefficient, the probe relaxes to the potential minimum of the linker before the next motor jump occurs, whereas the large probe cannot relax \cite{lade11}. Large ATP concentrations induce many forward and succesive backward motor jumps that are absent at lower ATP concentrations. These details are not captured in the coarse-grained trajectories.

\section{Motor models with several internal states}\label{sec:multistate}
\subsection{Explicit motor-bead dynamics and coarse-graining procedure}
\label{sec:CG-mult}
\begin{figure}[t]
 \centering
\includegraphics[width=1.0\linewidth]{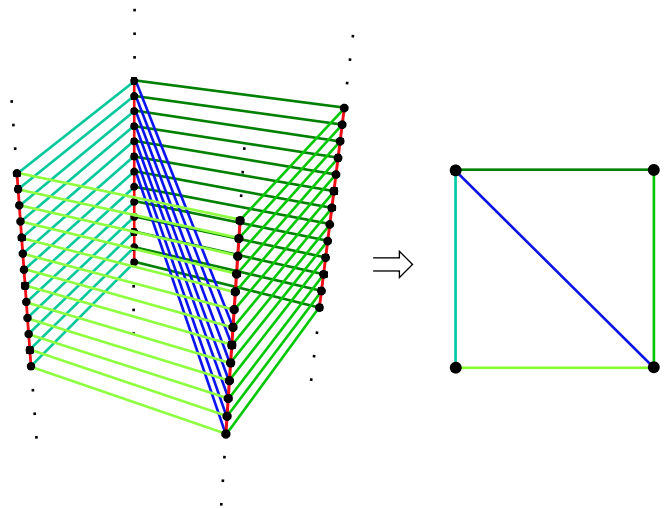}
\caption{(Color online) Network representation of a motor-bead model with four internal motor states and discretized state space of the probe particle (left). Each row of black dots represents one motor state while the dots themselves represent specific distances $y$ accessible to the probe particle (via the vertical red lines) within the same motor state. Transitions between motor states either leave $y$ the same (horizontal green lines) or can advance the motor by $d^{\alpha}_{ij}$ and change $y$ (diagonal blue lines).
The top view of this network corresponds to the coarse-grained version of this model (right).}
\label{fig:multi}
\end{figure}

In this section, we will generalize the model taking into account several different internal states of the motor labelled by $i$. The motor states represent the nodes and the transitions the edges of a network. Transitions between the motor states $i$ and $j$ change the free energy by 
\begin{align}
 \Delta F^{\alpha}_{ij}\equiv F_j-F_i-\Delta\mu_{ij}^{\alpha}
\end{align}
where $F_j-F_i$ is the free energy difference of the internal states of the motor and $\Delta\mu^{\alpha}_{ij}=-\Delta\mu^{\alpha}_{ji}$ is the free energy change of the solvent. Depending on the transition, $\Delta\mu^{\alpha}_{ij}$ is given by $\mu_\T$, $\mu_\D$, $\mu_\Ph$ or any combination thereof or $0$. Transitions may also advance the motor a distance $d^{\alpha}_{ij}=-d^{\alpha}_{ji}$.  Since we allow for several transitions connecting two states, we assign an additional index $\alpha$ to the transitions indicating which link between $i$ and $j$ is used. An example for the network of a full system comprising motor and probe particle is shown in Fig. \ref{fig:multi}, where the state space of the probe is discretized for better presentation.

The Fokker-Planck-type equation for such models is given by
\begin{align}
 \partial_t& p_i(y) = \partial_y\left( \left(\partial_y\,V(y)-\fex\right)p_i(y)+\partial_yp_i(y)\right)/\gamma \nonumber \\ 
	  & +  \sum_{j,\alpha} w^{\alpha}_{ji}(y+d^{\alpha}_{ij})p_j(y+d^{\alpha}_{ij})-w^{\alpha}_{ij}(y)p_i(y). \label{FPij}
\end{align} 
with transition rates of the motor that obey a LDB condition
\begin{align}
 \frac{w^{\alpha}_{ij}(y)}{w^{\alpha}_{ji}(y+d^{\alpha}_{ij})}=\exp[-\DF_{ij}-V(y+d^{\alpha}_{ij})+V(y)]. \label{eq:ldbmulti}
\end{align}

The coarse-grained version of such a model should take into account the different states of the motor as well as the several possible $\alpha$-transitions between $i$ and $j$. Thus, the motor network (including all motor cycles) should be conserved under coarse-graining.
To account for the several internal states, we require that the coarse-grained rates should obey a LDB condition and the operational current \cite{hill} from motor state $i$ to motor state $j$ via edge $\alpha$ should be conserved. The operational current is the sum over all $y$-dependent net transition currents that contribute to the transition $i\rightarrow j$. Conserving the operational currents corresponds to the condition of reproducing the correct mean velocity for the one-state model. The above conditions read

\begin{align}
 \frac{\Omega^{\alpha}_{ij}}{\Omega^{\alpha}_{ji}}=\exp[-\DF_{ij}-\fex d^{\alpha}_{ij}] \label{eq:CGldbmulti}
\end{align}
and
\begin{align}
 P_i\Omega^{\alpha}_{ij}-P_j\Omega^{\alpha}_{ji} = j^{\alpha}_{ij} \label{occ}
\end{align}
with the operational current
\begin{align}
  j^{\alpha}_{ij}&\equiv\int\limits_{-\infty}^{\infty} \left[p_i(y) w^{\alpha}_{ij}(y)-p_j(y+d^{\alpha}_{ij}) w^{\alpha}_{ji}(y+d^{\alpha}_{ij})\right] \mathrm{d}y \nonumber \\ &=-j^{\alpha}_{ji} \label{oc}
\end{align}
and the marginal distribution 
\begin{align}
 P_i=\int_{-\infty}^{\infty} p_i(y) \, \mathrm{d}y.
\end{align}
These equations can be solved for $\Omega^{\alpha}_{ij}$ and $\Omega^{\alpha}_{ji}$ using simple algebra which yields the rates
\begin{align}
  \Omega^{\alpha}_{ij}&=j^{\alpha}_{ij}\,\frac{\exp[-\DF_{ij}-\fex d^{\alpha}_{ij}]}{P_i\exp[-\DF_{ij}-\fex d^{\alpha}_{ij}]-P_j} \label{OPmulti}\\ 
  \Omega^{\alpha}_{ji}&=j^{\alpha}_{ij}\,\frac{1}{P_i\exp[-\DF_{ij}-\fex d^{\alpha}_{ij}]-P_j}. \label{OMmulti}
\end{align}
In principle, it is sufficient to use only eq. (\ref{OPmulti}), since $\Omega^{\alpha}_{ji}$ takes exactly this form with $j_{ij}^{\alpha}=-j_{ji}^{\alpha}, \Delta F^{\alpha}_{ij}=-\Delta F^{\alpha}_{ji}$ and $d_{ij}^{\alpha}=-d_{ji}^{\alpha}$. This equivalent procedure would be more symmetric and treat all transition rates on an equal footing but the LDB condition is then less obvious.
Note that without the LDB condition (\ref{eq:CGldbmulti}), the stated conditions of $P_i$ and $j^{\alpha}_{ij}$ would also be compatible with coarse-grained rates like the ones in, e.g., \cite{sant11,espo12}.

Transitions whose rates are independent of the linker elongation $y$ and hence have $d^{\alpha}_{ij}=0$ retrieve their original rate constants through this coarse-graining procedure. For such a transition, $j^{\alpha}_{ij}$ is given by
\begin{align}
 j^{\alpha}_{ij}&=P_iw^{\alpha}_{ij}-P_jw^{\alpha}_{ji}
\end{align}
with rates fulfilling the LDB condition $w^{\alpha}_{ij}/w^{\alpha}_{ji}=\exp[-\DF_{ij}]$. Inserting $j^{\alpha}_{ij}$ in eqs. (\ref{OPmulti}, \ref{OMmulti}) and using the LDB condition and $d^{\alpha}_{ij}=0$  immediately yields
\begin{align}
 \Omega^{\alpha}_{ij}=w^{\alpha}_{ij}, \quad \Omega^{\alpha}_{ji}=w^{\alpha}_{ji}.
\end{align}

Transitions with rates depending on $y$ but with $d^{\alpha}_{ij}=0$ have coarse-grained rates that depend on $\fex$ only implicitely via $j^{\alpha}_{ij}$ and $P_{i,j}$ as will be discussed below in section \ref{sec:kin} for the chemical transition rates of kinesin.

The rates determined from the LDB condition eq. (\ref{eq:CGldbmulti}), the populations $P_i$ and the operational currents are algebraically consistent with the fact that a full set of rates $\Omega^{\alpha}_{ij}$ will uniquely determine the populations $P_i$ on the coarse-grained network. Consistency can be seen by integrating the Fokker-Planck equation (\ref{FPij}) over $y$ yielding the coarse-grained master equation
\begin{align}
 \partial_t P_i=\sum_{j,\alpha}j_{ji}^{\alpha}=\sum_{j,\alpha} P_j\Omega_{ji}^{\alpha}-P_i\Omega_{ij}^{\alpha}, \label{FPCG}
\end{align}
whose stationary solution in the NESS can be expressed as a function of the rates $\Omega^{\alpha}_{ij}$ \cite{hill,hill66}. Thus, the expression of any current observable in terms of the operational currents is consistent with its expression in terms of cycle currents on the coarse-grained network.

\subsection{Time-scale separation}
Similar to the one-state-model, we explore the consequences of a putative time-scale separation between the dynamics of motor and probe for each motor transition. In the limit $\gamma\rightarrow 0$ (formally equivalent to $\varepsilon\rightarrow 0$ but here one would have several $\varepsilon_{ij}$ within the Fokker-Planck equation and all go to 0) the solution of eq. (\ref{FPij}) in the NESS becomes, analogously to \cite{raha07,bo14},
\begin{align}
 \hat{p}_{i}^{s}(y)=\hat{P}_{i}\exp[-V(y)+\fex y]/\mathcal{N}.
\end{align}
The marginal distribution can be obtained using eq. (\ref{FPij}) with its solution for fast bead relaxation
\begin{align}
 &\partial_t \hat{P}_{i}=\int_{-\infty}^{\infty} \partial_t \hat{p}_{i}^{s}(y)\,\mathrm{d}y \nonumber \\
&=\sum_{j,\alpha} \left(\hat{P}_{j} \langle w^{\alpha}_{ji}\rangle_y -\hat{P}_{i}\langle w^{\alpha}_{ij}\rangle_y\right) =\sum_{j,\alpha} \hat{j}^{\alpha}_{ji}=0. \label{Pi0th}
\end{align}
For Kramers-type transition rates like eqs. (\ref{wplus}, \ref{wminus}),
\begin{align}
 w^{\alpha}_{ij}(y)=&k^{\alpha}_{ij} \exp[\mu_{ij}^{\alpha,+}-V(y+d^{\alpha}_{ij}\tpa_{ij})+V(y)] \label{wplmulti}\\
 w^{\alpha}_{ji}(y)=&k_{ji}^{\alpha} \exp[\mu_{ij}^{\alpha,-}-V(y-d^{\alpha}_{ij}\tma_{ij})+V(y)],\label{wmimulti}
\end{align}
with $\mu^{\alpha,+}_{ij}-\mu^{\alpha,-}_{ij}=\Delta\mu^{\alpha}_{ij}$ and $k^{\alpha}_{ij}/k^{\alpha}_{ji}=\exp[-F_j+F_i]$, the $y$-averaged rates $\langle w^{\alpha}_{ij}\rangle_y$ and $\langle w^{\alpha}_{ji}\rangle_y$ become
\begin{align}
 \langle w^{\alpha}_{ij}\rangle_y=&k_{ij}^{\alpha}\exp[\mu_{ij}^{\alpha,+}-\fex d^{\alpha}_{ij}\tpa_{ij}] \\
 \langle w^{\alpha}_{ji}\rangle_y=&k_{ji}^{\alpha}\exp[\mu_{ij}^{\alpha,-}+\fex d^{\alpha}_{ij}\tma_{ij}].
\end{align}
The change of chemical free energy $\Delta\mu^{\alpha}_{ij}$ is split into $\mu^{\alpha,+}_{ij}$ and $\mu^{\alpha,-}_{ij}$ indicating that both directions of the transition can involve binding and release of the chemical species that account for $\Delta\mu^{\alpha}_{ij}$. The free energy change arising from changing the motor state, $F_j-F_i$, is incorporated in the attempt frequencies $k^{\alpha}_{ij}$ of the corresponding states.   
Inserting the operational current in the form of eq. (\ref{Pi0th}) with these averaged rates, simple calculus shows that the coarse-grained rates (\ref{OPmulti}) and (\ref{OMmulti}) reduce to
\begin{align}
 \hat{\Omega}^{\alpha}_{ij}=& k^{\alpha}_{ij}\exp[\mu_{ij}^{\alpha,+}-\fex d^{\alpha}_{ij}\tpa_{ij}]  \\
\hat{\Omega}^{\alpha}_{ji}=& k^{\alpha}_{ji}\exp[\mu_{ij}^{\alpha,-}+\fex d^{\alpha}_{ij}\tma_{ij}]
\end{align}
which is again consistent with transition rates of one-particle models that assume a purely exponential dependence on the external force.

\subsection{Example: $\text{F}_{\boldsymbol 1}$-ATPase with intermediate step}
\label{F19030}
\subsubsection{With external force}
\begin{figure}[b]
 \includegraphics[width=1.0\linewidth]{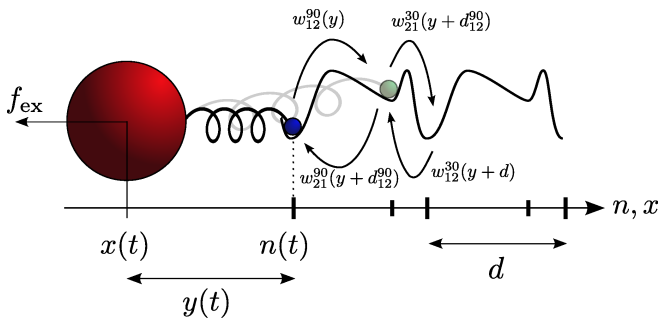}
\caption{(Color online) Schematic representation of a motor-bead model for the $\text{F}_{1}$-ATPase with two inernal states of the motor, 1 (blue (dark gray)) and 2 (pale green (light gray)). Transition between states 1 to 2 corresponding to the $90^{\circ}$ ($30^{\circ}$) substep are labelled with superscript $90$ ($30$). The transition rates are chosen accordingly to eqs. (\ref{wplmulti}, \ref{wmimulti}).}
\label{fig:9030}
\end{figure}

The $120^{\circ}$ step of the $\text{F}_{1}$-ATPase is known to consist of a $90^{\circ}$ and a $30^{\circ}$ substep \cite{yasu01}. Such a stepping behavior can be modelled with a unicyclic motor with two internal states. A schematic representation of a system comprising a probe particle and a motor with two internal states is shown in Fig. \ref{fig:9030}. The two different pathways for transitions between the states 1 and 2 correspond to the $90^{\circ}$ and $30^{\circ}$ substeps of the $\text{F}_{1}$-ATPase, respectively.

\begin{figure}[t]
\hfill \subfigure{\includegraphics[width=0.99\linewidth]{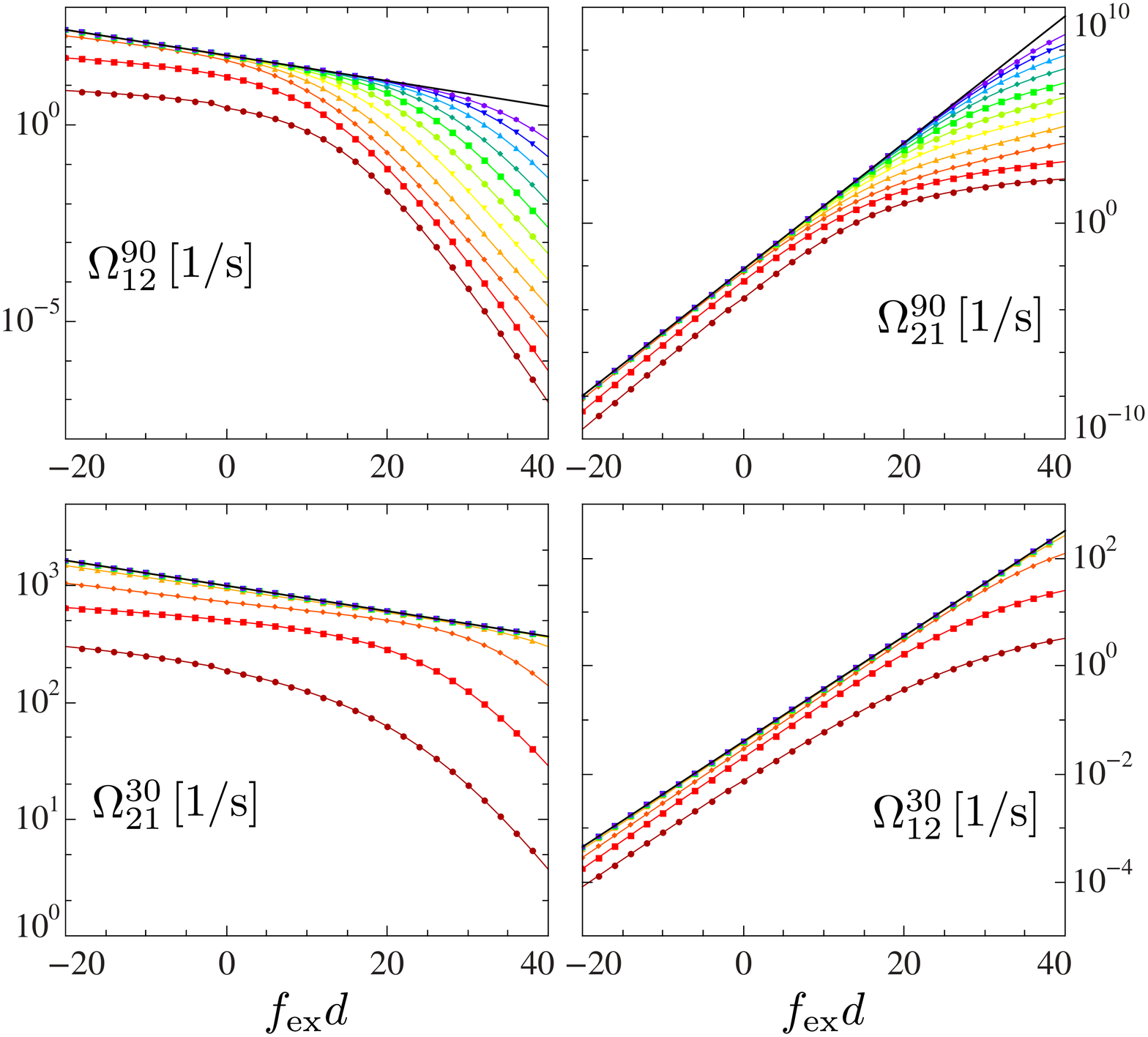}}
\raggedright \subfigure{\includegraphics[width=0.95\linewidth]{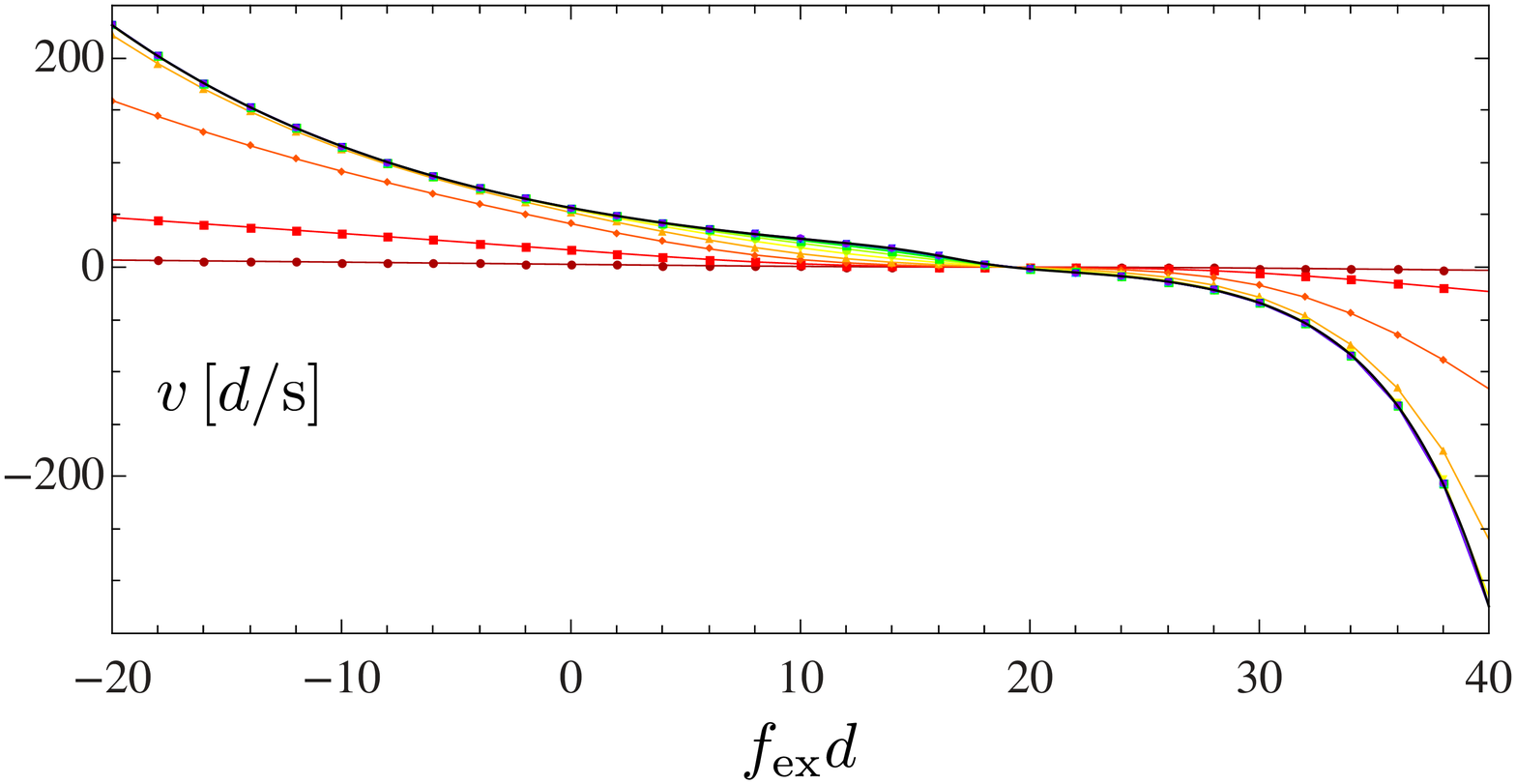}}
\caption{(Color online) Coarse-grained rates for the $90^{\circ}$ (top) and the $30^{\circ}$ (center) substep and average velocity (bottom) as functions of $\fex d$ for various $\gamma$ in the range $5\text{s}/d^2  \geq \gamma  \geq 5\cdot 10^{-10}\text{s}/d^2$ (from bottom to top). With decreasing $\gamma$, the rates and the velocity approach their corresponding fast-bead limit (solid black lines). Parameters: $\kappa=40 d^{-2}$, $c_\T=c_\D=2\cdot10^{-6}$M, $c_\Ph=10^{-3}$M, $\tp_{90,30}=0.1$, $k_{12}^{90}\exp[\mu_\T^{\text{eq}}]/c_\T^{\text{eq}}=3\cdot 10^{7}(\text{Ms})^{-1}$, $k_{21}^{90}\exp[\mu_D^{\text{eq}}]/c_\D^{\text{eq}}=3667.5(\text{Ms})^{-1}$, $k_{21}^{30}=1000\text{s}^{-1}$, $k_{12}^{30}\exp[\mu_P^{\text{eq}}]/c_\Ph^{\text{eq}}=40(\text{Ms})^{-1}$. The attempt frequencies are chosen on the basis of \cite{yasu01,adac07} where very small probe particles have been used.}
\label{OPM9030} 
\end{figure}

Like in section \ref{sec:F1120} for the one-state model, we examine the coarse-grained rates for the $90^{\circ}$ and $30^{\circ}$ steps and the velocity which are shown in Fig. \ref{OPM9030}. Similar to the $120^{\circ}$-scenario, the rates approach their fast-bead limit with decreasing $\gamma$.

As in the one-step model, the dependence of the coarse-grained rates on the external force shows two regimes. For small external forces, the rates can be well approximated by a single exponential dependence on $\fex$ with slope $\pm d^{\alpha}_{ij}\theta^{\alpha,\pm}_{ij}$ in most cases. For large probe particles, however, the rates neither match the absolute value nor show mono-exponential dependence on $\fex$ with the above slope. For large forces, the forward rates decay faster whereas the backward rates grow more slowly than in the fast-bead limit. 

Concerning the average velocity, strong deviations from the fast-bead limit occur only for the largest friction coefficients. Using small beads, the force-velocity relation resulting from our coarse-graining procedure coincides well with the one obtained from a one-particle model due to the fact that the velocity involves only differences of the rates multiplied with the marginal distribution rather than the rates themselves. For large external forces and small $\gamma$, the velocity is significantly smaller than in the one-state model since the motor has to take two successive steps to cover the full $d$. The force-velocity relations for the two-state as well as for the one-state model reproduce very well the exerimentally determined force-velocity relation from \cite{toya11} for the corresponding value of the friction coefficient $\gamma$.

The limiting cases $\fex\rightarrow\pm\infty$ are more involved here than in the one-state model since one has to account for the dependence of the $P_i$'s on the external force. However, as long as the $P_j$'s do not decay faster than $\exp[-\fex d^{\alpha}_{ij}]$, it is still possible to approximate the rates (\ref{OPmulti}, \ref{OMmulti}) by
\begin{align}
 \Omega^{\alpha}_{ij}&\approx -j^{\alpha}_{ij}\exp[-\DF_{ij}-\fex d^{\alpha}_{ij}]/P_j, \\
 \Omega^{\alpha}_{ji}&\approx -j^{\alpha}_{ij}/P_j 
\end{align}
since $P_i$ is bounded by $1$.

For the $\text{F}_{1}$-ATPase model, the numerical analysis in the $\fex\rightarrow\infty$ limit yields a linear dependence of $\av{y}$ and $j_{ij}^{\alpha}$ on $\fex$.  We also find that $P_{2}$ decays exponentially while $P_1$ approaches $1$. Hence, $\Omega^{90}_{12}$ and $\Omega^{30}_{21}$ decay exponentially with slope $-d^{90}_{12}=-0.75d$ and $-d^{30}_{21}=-0.25d$, respectively, like in the one-state model but $\Omega_{21}^{90}$ now grows exponentially with a smaller exponent while $\Omega_{12}^{30}$ still grows linearly.

\subsubsection{Without external force}

Just as for the one-state-model, we examine the dependence of the coarse-grained rates on the ATP concentration in the absence of external forces.

\begin{figure}[t]
\hfill \subfigure{\includegraphics[width=0.99\linewidth]{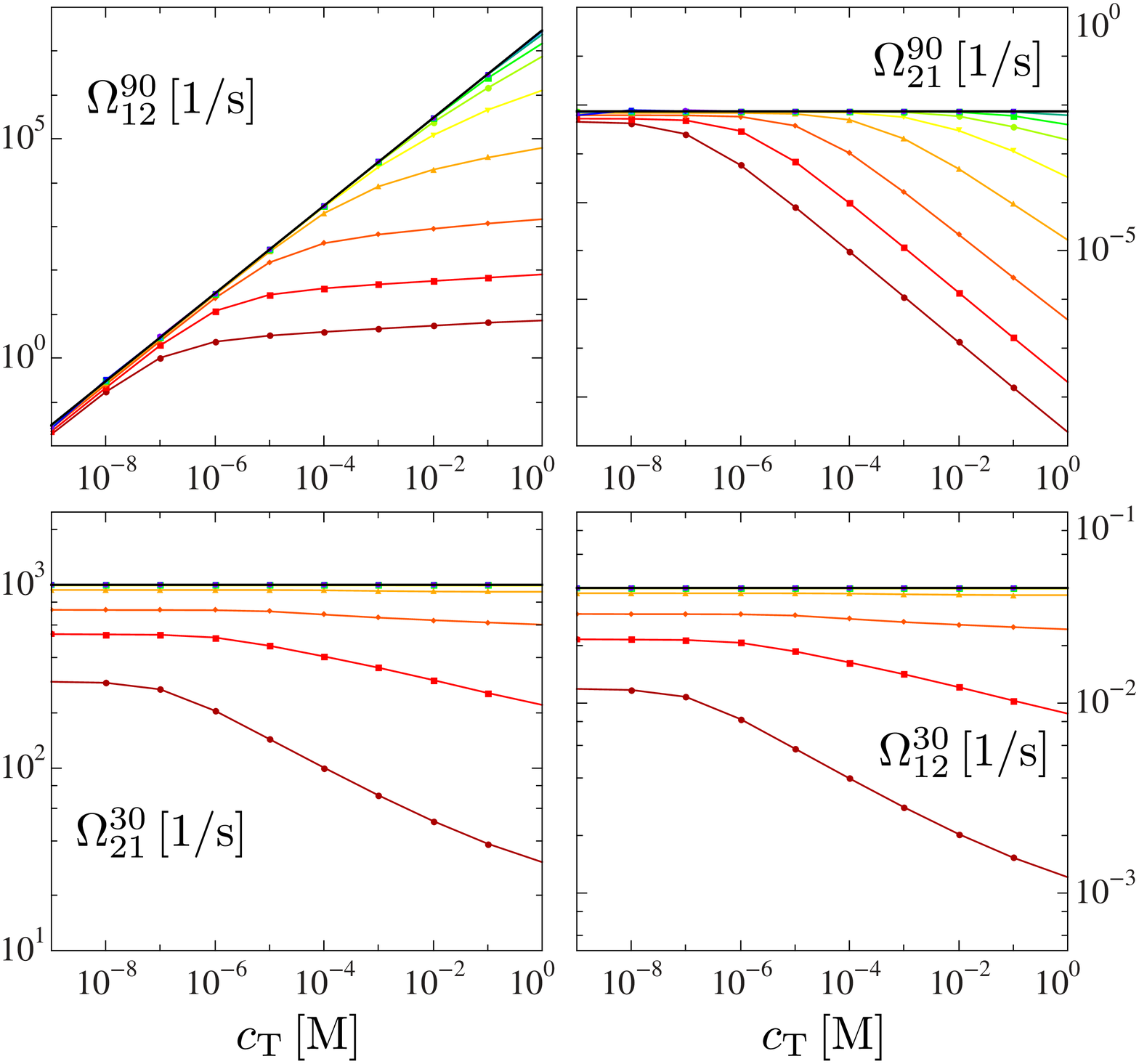}}
\raggedright \subfigure{\includegraphics[width=0.965\linewidth]{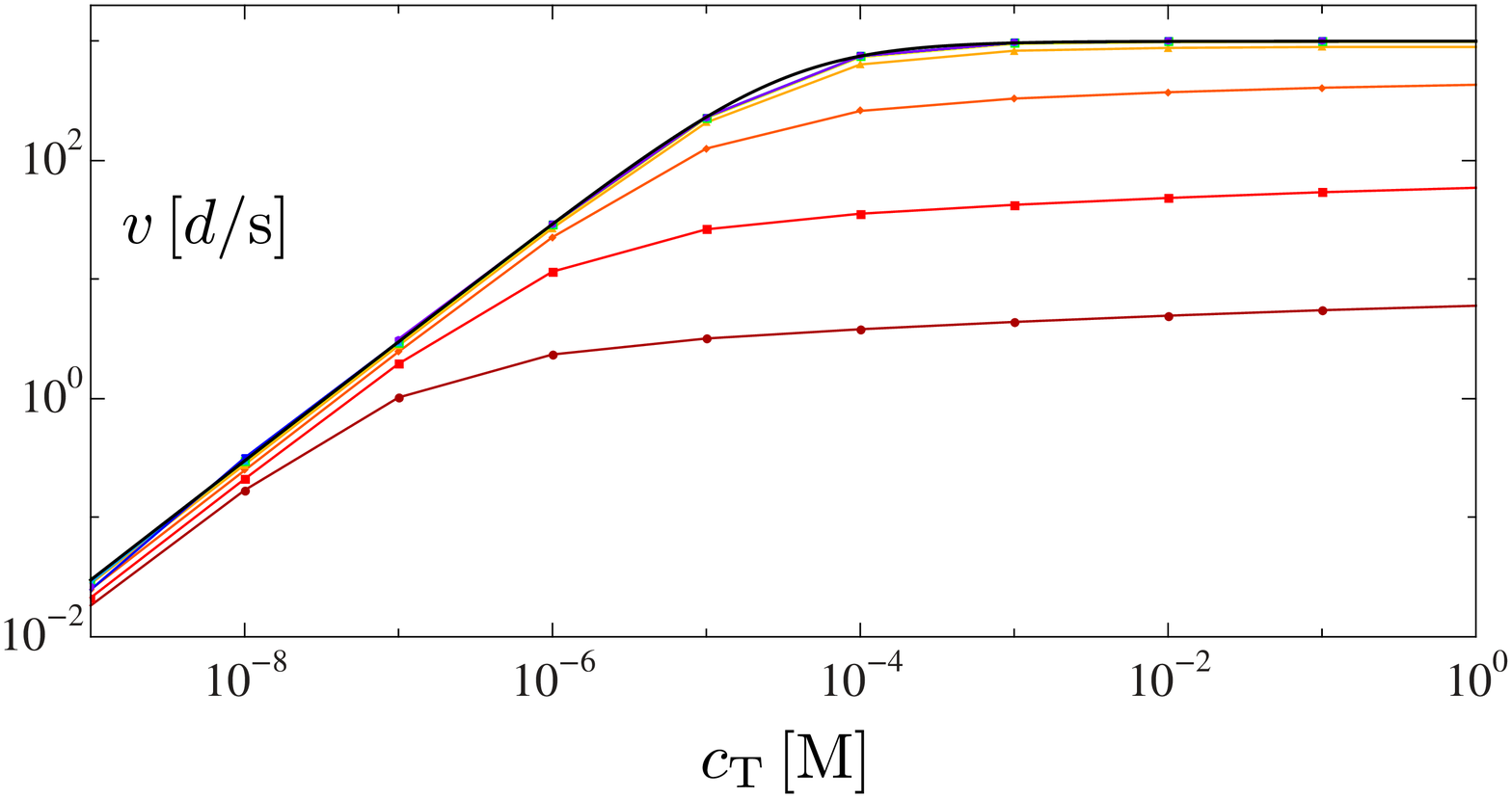}}
\caption{(Color online) Coarse-grained rates for the $90^{\circ}$ (top) and the $30^{\circ}$ (center) substep and average velocity (bottom) for various $\gamma$ and $\fex=0$ as functions of $c_\T$. Since $c_\D$ and $c_\Ph$ are fixed, $\Delta\mu$ also increases with $c_\T$. The rates and the velocity approach the fast-bead approximation (solid black lines). Parameters: $c_\D=2\cdot10^{-6}\text{M}$, $c_\Ph=1\cdot10^{-3}\text{M}$, $\kappa=40 d^{-2}$, $\tp_{90,30}=0.1$, $\gamma$ in the range $5\,\text{s}/d^2 \, \geq \gamma \, \geq \,\,5\cdot 10^{-10}\,\text{s}/d^2$ (from bottom to top). }
\label{fig:gamma9030}
\end{figure}

Fig. \ref{fig:gamma9030} shows the coarse-grained rates for the $90^{\circ}$ and the $30^{\circ}$ substeps as well as the avergage velocity. With decreasing $\gamma$, the coarse-grained rates approach the mass action law kinetics for the corresponding one-particle rates. In contrast to the one-state model, even in this limit, the velocity shows saturation. This is due to the fact that the timescale of the hydrolysis reaction is independent of the ATP concentration and represents the limiting effect for the velocity. The dependence of the average velocity on the ATP concentration is reminiscent of a Michaelis-Menten kinetics and coincides well with experimenal results for several different probe particles as shown in \cite{yasu01}.

For large beads, the coarse-graining process yields rates that are no longer linear in the corresponding concentrations. In this regime, the sub-linear dependence of the velocity on the ATP concentration appears already for smaller ATP concentrations. Comparing the velocity curves of the two-state model with the one-state model, we find that for large beads the velocity curves almost coincide since in this regime the limiting effect for the velocity is the friction experienced by the bead. Thus, using large probe particles it is not possible to infer the underlying motor dynamics from the characteristics of the velocity as a function of the ATP concentration \cite{lade11}.

\begin{figure}[t]
 \centering
\includegraphics[width=1.0\linewidth]{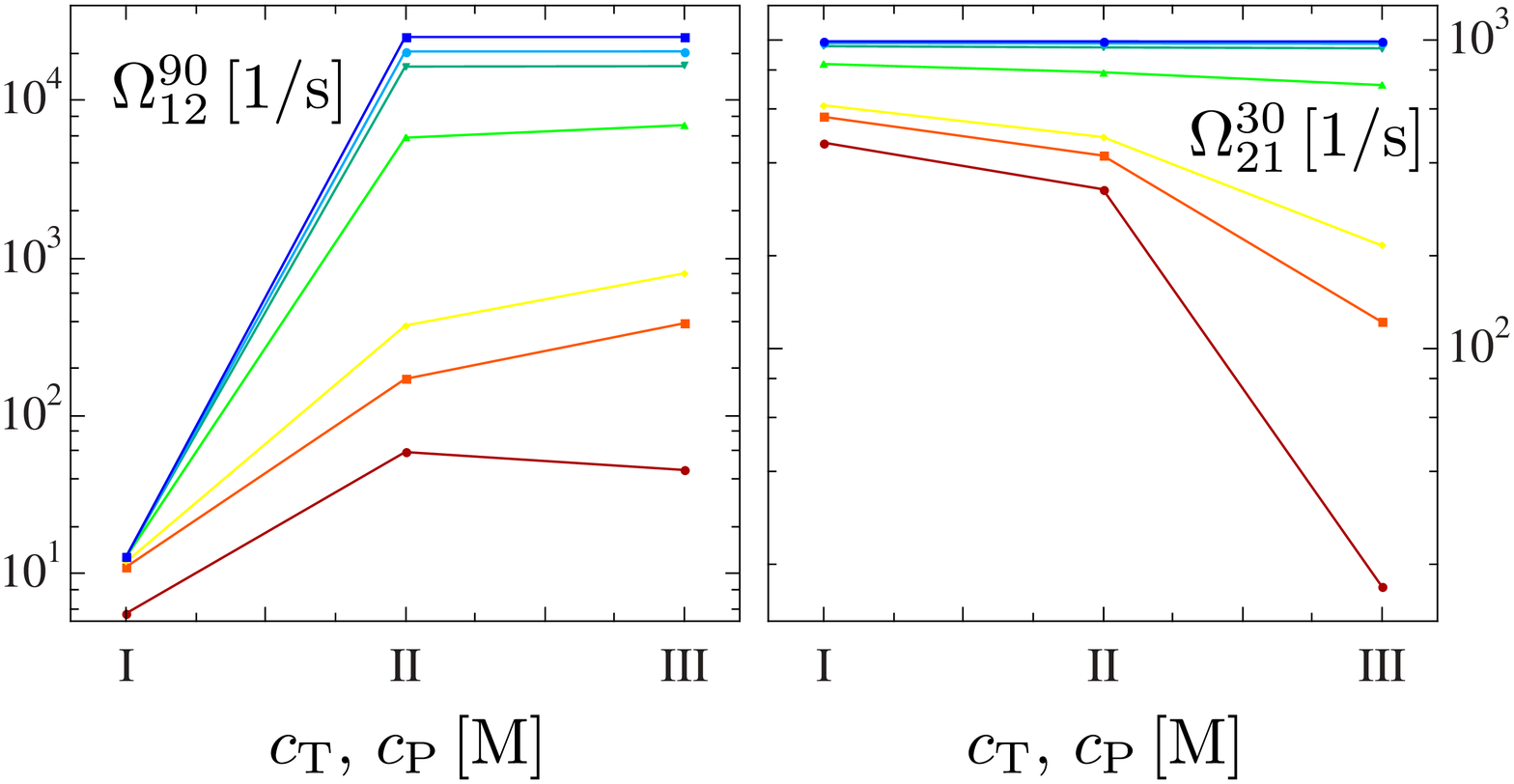}
\caption{(Color online) Coarse-grained forward rates for various $\gamma=2.1\text{s}/d^2$, $0.26\text{s}/d^2$, $0.14\text{s}/d^2$, $0.017\text{s}/d^2$, $0.003\text{s}/d^2$, $0.001\text{s}/d^2$, $3.8\cdot 10^{-4}\text{s}/d^2$ (from bottom to top) for the parameter sets I: $c_\T=430$ nM, $c_\Ph=1$ nM; II: $c_\T=1$ mM, $c_\Ph=1$ nM; III: $c_\T=1$ mM, $c_\Ph=200$ mM as used in \cite{wata13}. Parameters that are the same for all sets I-III: $c_\D=1$ nM, $\kappa=40 d^{-2}$, $\tp_{90,30}=0.1$ and the attempt frequencies $k_{ij}^{\alpha}$ as given in Fig \ref{OPM9030}. The values of $c_\D=c_\Ph=1$ nM are a rough estimate because there is no information about these concentrations in \cite{wata13}.}
\label{wata}
\end{figure}

Fig. \ref{wata} shows the coarse-grained forward rates for three different nucleotide concentrations and for various $\gamma$ chosen as in the experiment \cite{wata13}. We find that the $90^{\circ}$ rate depends only weakly on $\gamma$ for small ATP concentrations which is reminiscent of the experimental observation that the ATP binding rate to the motor depends only weakly on the size of the probe \cite{wata13}. However, for large ATP concentrations that were not investigated in the experiment, the $90^{\circ}$ rate shows a strong dependence on $\gamma$. This is due to the fact that for small ATP concentrations the relaxation times of all probe particles are in the order of, or even faster than, the motor jump rates. The results for the $30^{\circ}$ rate are consistent with experimental results for the hydrolysis rate \cite{wata13}. Increasing $c_\Ph$ decreases the $\text{P}_{\text{i}}$ release rate in the experiment as it decreases the $30^{\circ}$ rate here.

\subsection{Example: Kinesin}
\label{sec:kin}

\begin{figure}[h]
 \centering
\includegraphics[width=0.6\linewidth]{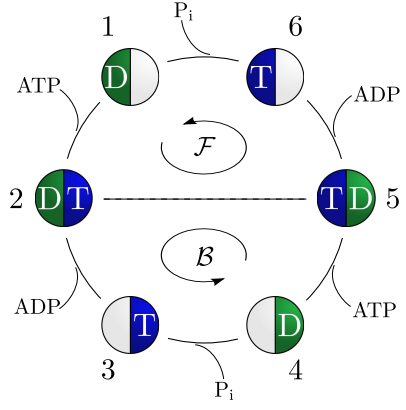}
\caption{(Color online) 6-state-model representing a kinesin motor adapted from \cite{liep07a}. The transition between states $2$ and $5$ is purely mechanical and corresponds to a step of length $d$ whereas all other transitions are pure chemical transitions. The motor model includes three cycles: $\mathcal{F}$ which, in the $+$ direction, includes ATP hydrolysis and forward stepping, $\mathcal{B}$ which includes ATP hydrolysis and backward stepping in its $+$ direction and a pure chemical cycle (around the circle) that includes hydrolysis/synthesis of two ATP.}
\label{Kinesin_LL}
\end{figure}

As a final more complex example, we apply our coarse-graining method to a model with a multi-state motor. We choose the well-studied 6-state-model representing a kinesin motor introduced in \cite{liep07a}, see Fig. \ref{Kinesin_LL}. Implementing the probe particle and an elastic linker $V(y)$, we adopt the transition rates of the motor from \cite{liep07a} and replace the dependence on the external force by the dependence on the elongation of the linker, 
\begin{align}
 &\wpl_{25}(y)=k_{25}\exp[-V(y+d\tp)+V(y)], \\
&\wmi_{52}(y)=k_{52}\exp[-V(y-d\tm)+V(y)], \\
&\wpl_{ij,\text{chem}}=k_{ij}\frac{2 \exp[\mu_{ij}^{+}]}{1+\exp[\partial_y V(y)\chi_{ij}]}, \label{wpch}\\
&\wmi_{ji,\text{chem}}=k_{ji}\frac{2 \exp[\mu_{ij}^{-}]}{1+\exp[\partial_y V(y)\chi_{ij}]}. \label{wmch}
\end{align}
The first two rates belong to the mechanical transition, the lower two rates represent the chemical transitions which depend on the instantaneous force exerted by the linker with a chemical load-sharing factor $\chi_{ij}$, see \cite{liep07a}. The change of chemical free energy $\mu_{ij}^{\pm}=\mu_\T,\mu_\D,\mu_\Ph$ depends on which transition involves binding of the corresponding nucleotide. We choose again $V(y)=\kappa y^2/2$.

\begin{figure}[h]
 \centering
\includegraphics[width=1.0\linewidth]{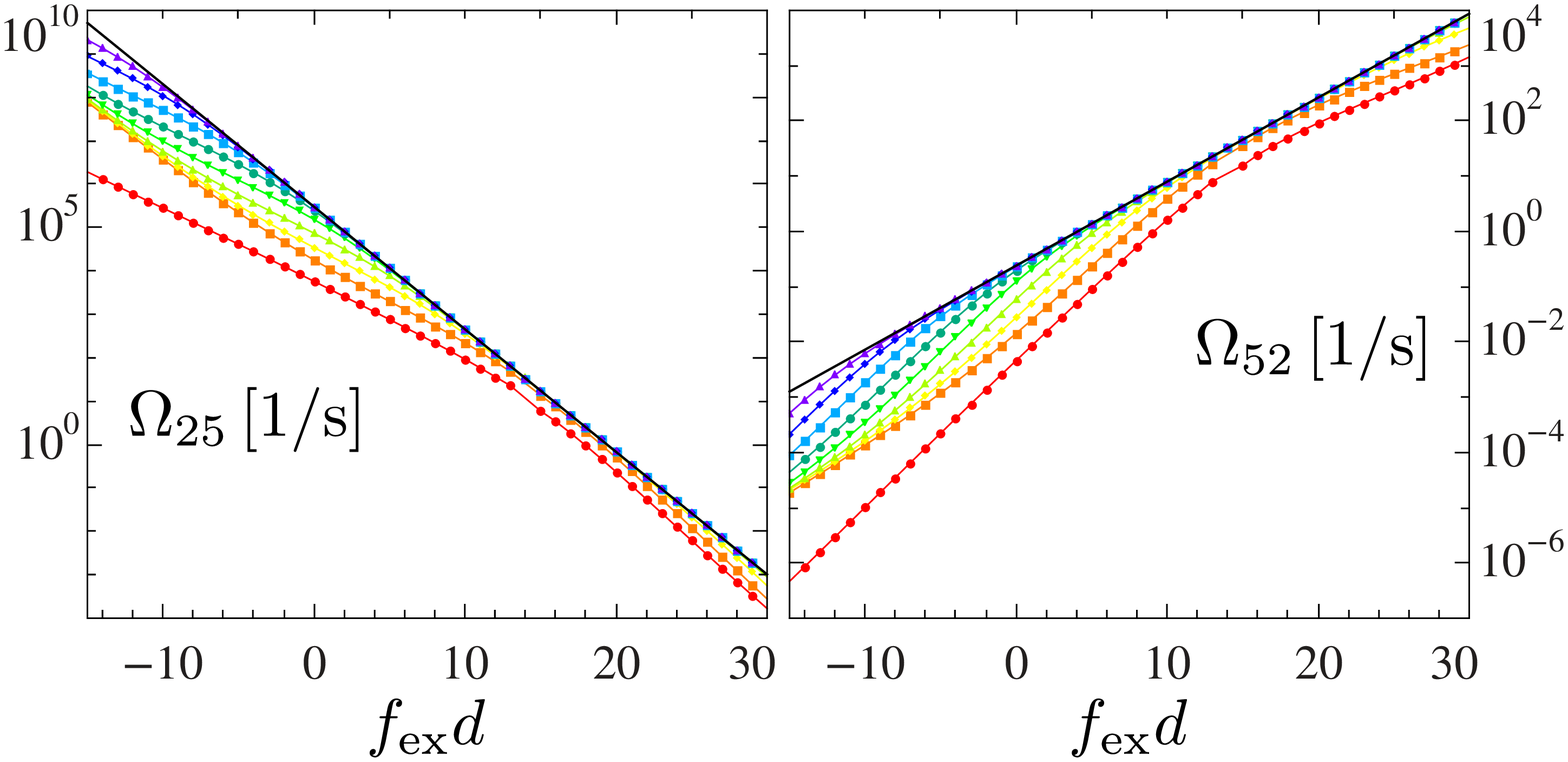}
\caption{(Color online) Coarse-grained rates for the mechanical transitions (with $y$-dependence) for various $\gamma$ in the range $0.077\text{s}/d^2  \geq \gamma  \geq 7.7\cdot 10^{-10}\text{s}/d^2$ (from bottom to top). The rates approach the one-particle rates from \cite{liep07a} (solid black lines). Parameters: $\kappa=10 d^{-2}$ \cite{cart05}, $c_\T=0.001$M, $c_\D=c_\Ph=10^{-9}$M (estimated), $\tp=0.65$, $\chi_{ij}=0.25,0.15$, $k_{12}\exp[\mu_T^{\text{eq}}]/c_\T^{\text{eq}}=k_{45}\exp[\mu_T^{\text{eq}}]/c_\T^{\text{eq}}=2\cdot 10^{6}(\text{Ms})^{-1}$, $k_{21}=k_{23}=k_{34}=k_{56}=k_{61}=100(\text{s})^{-1}$, $k_{32}\exp[\mu_D^{\text{eq}}]/c_\D^{\text{eq}}=k_{65}\exp[\mu_D^{\text{eq}}]/c_\D^{\text{eq}}=2\cdot 10^{4}(\text{Ms})^{-1}$, $k_{43}\exp[\mu_\Ph^{\text{eq}}]/c_\Ph^{\text{eq}}=k_{16}\exp[\mu_\Ph^{\text{eq}}]/c_\Ph^{\text{eq}}=2\cdot10^{4}(\text{Ms})^{-1}$, $k_{25}=3\cdot10^{5}(\text{s})^{-1}$, $k_{52}=0.24(\text{s})^{-1}$, $k_{54}=(k_{52}/k_{25})^2k_{21}$.}
\label{OPOMmech}
\end{figure}

\begin{figure}[t]
 \centering
\includegraphics[width=1.0\linewidth]{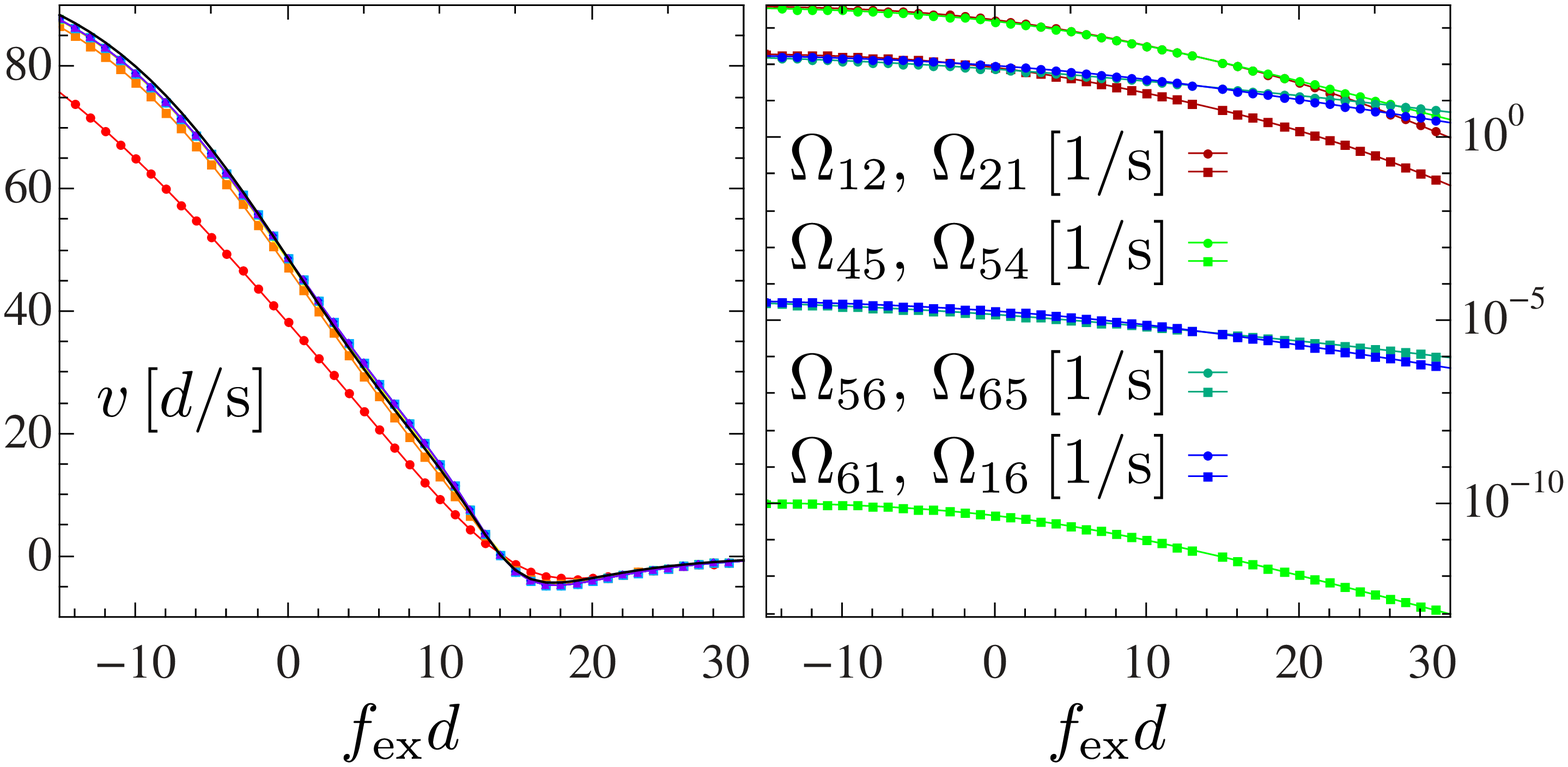}
\caption{(Color online) Left: Average velocity for the model with probe particle (colored lines with symbols) an the one-particle model from \cite{liep07a} (solid black line). Right: Coarse-grained rates for chemical transitions (with $y$-dependence) for $\gamma=0.077 \text{s}/d^2$. Other parameters as given in Fig. \ref{OPOMmech}.}
\label{OPOMch}
\end{figure}

The coarse-grained rates for the mechanical transition are shown in Fig. \ref{OPOMmech}. With decreasing $\gamma$, the rates approach their fast-bead limit which corresponds to the rates used in \cite{liep07a} while strong deviations occur for finite $\gamma$ especially for assisting external forces. The friction coefficient of a probe of size $500$ nm as in \cite{cart05} can be calculated using Stokes' law yielding $\gamma\simeq 7.7\cdot 10^{-5}d^2/$s. For friction coefficients in this range (light green (light gray) line with triangles), our coarse-grained rates show a distinct deviation from the one-particle rates (solid black lines). However, the average velocity (obtained from our coarse-grained rates) as function of the external force  coincides very well for almost all $\gamma$ with the velocity curve obtained from the bare motor model, see Fig. \ref{OPOMch}. Like for the $\text{F}_{1}$-ATPase model discussed in section \ref{F19030}, this agreement is due to the fact that the velocity involves only the difference of the rates multiplied with the marginal distribution. If one investigates only force-velocity curves, the discrepancies between the coarse-grained rates and the one-particle rates are hardly visible.

In contrast to the coarse-grained rates of the $\text{F}_{1}$-ATPase models, the coarse-grained rates for the mechanical transition of the kinesin model show more structure especially for negative, i.e., assisting external forces. Since the kinesin model contains several internal motor cycles, depending on the external force the dominant cycle can change leading to crossover regimes with changing weight of the probabilities $P_{i}$.

The dependence of the coarse-grained rates for chemical transitions on the external force is visible in Fig. \ref{OPOMch}. Although there is no explicit dependence on external forces for pure chemical rates since $d_{ij}^{\alpha}=0$, $j_{ij}^{\alpha}$ and $P_i$ depend on $\fex$ via $y$. The operational current for transitions within the $\mathcal{F}$-cycle in $+$ direction decreases with increasing $\fex$ whereas the operational currents within the $\mathcal{B}$-cycle in $+$ direction slightly increase with $\fex$ which can be explained intuitively since the motor prefers ``backward'' cycles for large opposing forces. However, all coarse-grained rates decrease with increasing $\fex$ similar to the bare motor rates (\ref{wpch}, \ref{wmch}) which decrease with larger $y$; a situation that is more likely to appear for large external forces.

\section{Experimental Implementation}
\label{sec:exp}

In order to practically apply the coarse-grained description, one has to determine the marginal distributions $P_i$, the operational currents $j_{ij}^{\alpha}$ and the free energy differences $\Delta F_{ij}^{\alpha}$. For multi-state motors, this is a rather challenging task since only a few quantities can be extracted reliably from the experimentally measured trajectory of the probe. Note, however, that this problem does not happen exclusively in our approach but is inevitable whatever method is used to infer motor properties from such trajectories. 

In the following, using the $90$-$30$ model for the $\text{F}_{1}$-ATPase, we will illustrate how these quantities can be estimated. If all motor transitions involve mechanical transitions with different step sizes, the plateaus in the probe trajectory can be assigned to specific corresponding motor states. Since after a large enough time interval all possible transitions will have occured, one is also able to reconstruct the links connecting the states. The marginal distributions $P_i$ are then given as the fraction of time that the corresponding motor state is occupied. 
For the two-state model of the $\text{F}_{1}$-ATPase, we assign plateaus in the probe trajectory that are followed by a fast $90^{\circ}$ forward or $30^{\circ}$ backward displacement to motor state $i=1$ and plateaus that are followed by fast $90^{\circ}$ backward or $30^{\circ}$ forward displacement to $i=2$. In principle, there are several possibilities to reconstruct hidden variables from partially visible trajectories \cite{cart08,muel10,litt11}. Here, we will use a simple algorithm which sets $i=2$ if four consecutive data points are within a specific range around $90^{\circ}$ and otherwise $i=1$. The marginal distributions $P_1$, $P_2$ are then represented by the fraction of data points with assigned $i=1,2$.

\begin{figure}[t]
 \centering
\includegraphics[width=1.0\linewidth]{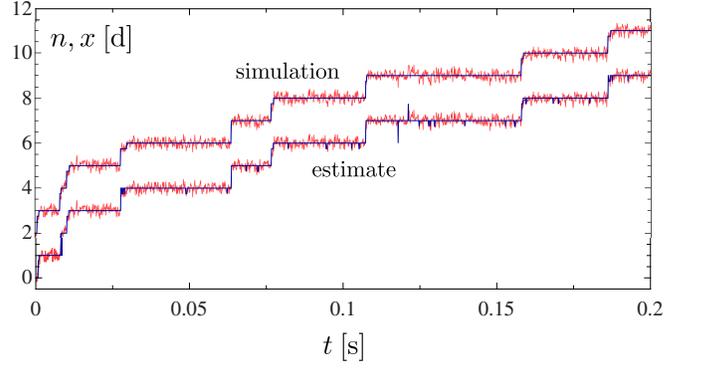}
\caption{(Color online) Comparison of the simulated trajectory of motor and probe with a trajectory of the probe and the estimated motor position that was reconstructed using the simulated trajectory of the probe. The trajectories are shifted for better visibility. Parameters: $\kappa=40 d^{-2}$, $\gamma=0.005\,\mathrm{s}/d^2$, $c_\T=c_\D=2\cdot10^{-6}$M, $c_\Ph=10^{-3}$M, $c_\T^{\text{eq}}=3.33\cdot10^{-7}$M, $c_\D^{\text{eq}}=0.0682$M, $c_\Ph^{\text{eq}}=1$M, $\fex=0$, lower boundary to set $i=2$: $x-\left\lfloor{x}\right\rfloor=0.375d$, upper boundary to set $i=2$: $x-\left\lfloor{x}\right\rfloor=0.89d$. }
\label{Inferred_motor}
\end{figure}

If the motor is not very complex, the operational currents $j_{ij}^{\alpha}$ can be obtained rather easily since they are precisely the net currents between two motor states. For unicyclic motors, all operational currents are equal to the average velocity divided by $d$, the operational current of an ATP binding transition is the net disappearance rate of ATP in the solution (given that there are no other ATP binding reactions), etc.. If all motor transitions involve mechanical transitions with different step sizes, the operational currents between any two states can be obtained by counting the number of transitions of a specific step size from $i\rightarrow j$, $n^{\alpha}_{ij}$, and $j\rightarrow i$, $n^{\alpha}_{ji}$. The (time) average of this current using one long trajectory of length $t_\text{tot}$ is then given by 
\begin{align}
 j_{ij}^{\alpha}=(n^{\alpha}_{ij}-n^{\alpha}_{ji})/t_\text{tot}.
\end{align}
In our example, in order to estimate $j_{12}^{90}$ we have to count the number of sudden displacements of ``size'' $90^{\circ}$ either from the trajectory of the probe directly or from the reconstructed trajectory of the motor using the assignment rule mentioned above. If the time resolution of the trajectory is very coarse or if the reconstruction method is rather inaccurate, jumps that consist of fast consecutive $90$ and $30$ jumps with apparent step size $120^{\circ}$ will appear which have to be included in the number of $90^{\circ}$ (and also $30^{\circ}$) jumps. Fig. \ref{Inferred_motor} shows a reconstructed motor trajectory obtained with the algorithm mentioned above. We have used a trajectory of the probe from our simulations as ``experimental data''. Compared to the original motor trajectory, this reconstruction captures the average dynamics quite well. Large fluctuations of the probe can generate additional apparent motor jumps in the reconstructed trajectory that are absent in the original one.

Finally, the estimation of the free energy difference $\Delta F_{ij}^{\alpha}=F_j-F_i-\Delta\mu^{\alpha}_{ij}$ is slightly more involved. In equilibrium ($\Delta\mu=0$, $\fex=0$), detailed balance holds, 
\begin{align}
 \frac{w^{\alpha}_{ij}(y)}{w^{\alpha}_{ji}(y+d^{\alpha}_{ij})}=\frac{p^{\text{eq}}_j(y+d^{\alpha}_{ij})}{p^{\text{eq}}_i(y)},
\end{align}
with the Boltzmann distribution $p^{\text{eq}}_i(y)=P_i^{\text{eq}}\exp[-V(y)]/\mathcal{N}$. Inserting this expression  yields 
\begin{align}
P^{\text{eq}}_j/P^{\text{eq}}_i=\exp[-F_j+F_i+\Delta\mu^{\alpha,\text{eq}}_{ij}]\equiv \exp[-\Delta\mathcal{F}^{\alpha}_{ij}]
\end{align}
for the marginal distributions in equilibrium. Note that $\Delta\mu^{\alpha,\text{eq}}_{ij}\neq 0$ if the corresponding transition comprises only binding or release of nucleotides. Thus, the equilibrium free energy difference $\Delta\mathcal{F}^{\alpha}_{ij}$ (which explicitely depends on the equilibrium concentrations) can be obtained from the ratio of the marginal distributions under equilibrium conditions. Using $\mu_i=\mu_i^{\text{eq}}+\ln(c_i/c_i^{\text{eq}})$, we find that
\begin{align}
 \Delta F^{\alpha}_{ij}=\Delta\mathcal{F}^{\alpha}_{ij}\pm\sum_k \ln\frac{c_k}{c_k^{\text{eq}}}
\end{align}
with $k=\T,\D,\Ph$ and the sign depending on which binding or release event corresponds to the transition $ij,\alpha$ \cite{seif11}. Hence, the free energy difference $\Delta F^{\alpha}_{ij}$ needed for the coarse-grained rates can be expressed by the equilibrium free energy difference $\Delta\mathcal{F}^{\alpha}_{ij}$ obtained from experimental data at equilibrium conditions and the nucleotide concentrations with respect to the equilibrium concentrations corresponding to the conditions used to obtain $\Delta\mathcal{F}^{\alpha}_{ij}$. 
For the $90$-$30$ model, we have $P^{\text{eq}}_2/P^{\text{eq}}_1=\exp[-\Delta\mathcal{F}_{12}^{90}]=\exp[-\Delta\mathcal{F}_{12}^{30}]$ with $-\Delta\mathcal{F}_{12}^{90}=-F_2+F_1+\mu_\T^{\text{eq}}-\mu_\D^{\text{eq}}=-F_2+F_1+\mu_\Ph^{\text{eq}}=-\Delta\mathcal{F}_{12}^{30}$ since $\Delta\mu=0$ in equilibrium.

Once these quantities have been estimated, there are no additional fit parameters needed or left. All concentrations as well as the external force are usually known from the experimental setup. To obtain the coarse-grained rates from the probe trajectory of our $90$-$30$ model, we then proceed as follows. First, we choose equilibrium conditions and obtain $\Delta\mathcal{F}_{12}$ from the ratio of marginal distributions. Then, we change to non-equilibrium concentrations and estimate $P_1$, $P_2$ and the operational current $j_{12}^{90}$. The coarse-grained rates, according to eqs. (\ref{OPmulti}, \ref{OMmulti}), are then given by
\begin{align}
 \Omega^{90}_{12}&=j^{90}_{12}\frac{c_\T c_\D^{\text{eq}} \exp[-\Delta\mathcal{F}_{12}-\fex d_{12}^{90}]/(c_\T^{\text{eq}}c_\D)}{P_1c_\T c_\D^{\text{eq}} \exp[-\Delta\mathcal{F}_{12}-\fex d_{12}^{90}]/(c_\T^{\text{eq}}c_\D)-P_2}, \\
\Omega^{90}_{21}&=j^{90}_{12}\frac{1}{P_1c_\T c_\D^{\text{eq}}\exp[-\Delta\mathcal{F}_{12}-\fex d_{12}^{90}]/(c_\T^{\text{eq}}c_\D)-P_2},  \\
\Omega^{30}_{21}&=j^{90}_{12}\frac{c_\Ph^{\text{eq}} \exp[\Delta\mathcal{F}_{12}-\fex d_{21}^{30}]/c_\Ph}{P_2c_\Ph^{eq}\exp[\Delta\mathcal{F}_{12}-\fex d_{21}^{30}]/c_\Ph-P_1}, \\
\Omega_{12}^{30}&=j^{90}_{12}\frac{1}{P_2c_\Ph^{\text{eq}} \exp[\Delta\mathcal{F}_{12}-\fex d_{21}^{30}]/c_\Ph-P_1}.
\end{align}
A comparison of the coarse-grained rates and related quantities obtained from the full theoretical model and from the reconstructed one estimated using the probe trajectory is shown in table \ref{table_1}. 
We find quite good agreement between the original and the reconstructed quantities with a maximum error of $14\%$ except for the $\Omega^{30}_{ij}$ rates which have a maximum error of $24\%$.

The $90$-$30$ model thus provides a useful demonstration of the experimental applicability of the coarse-graining method showing that it is possible to estimate the coarse-grained rates from experimental accessible data if the underlying motor network is not too complex. Considering the simplicity of the applied reconstruction method, the accuracy of the estimates is rather encouraging.

\begin{table}
\caption{Comparison of the coarse-grained rates and other relevant quantities obtained from the simulation of the full model with the ones estimated using the reconstructed motor trajectory. The trajectory used to obtain these values is shown in Fig. \ref{Inferred_motor}.}
\label{table_1}
 \centering
\begin{tabular}{|c||c|c|}
 \hline
   & simulation & estimate \\ \hline\hline
$P_1$ & 0.944 & 0.952 \\ \hline
$P_2$ & 0.056 & 0.048 \\ \hline
$j_{12}^{90}$ [1/s] & 52.292 & 52.246 \\ \hline
$\Delta\mathcal{F}_{12}$ & 3.216 & 3.140 \\ \hline
$\Omega_{12}^{90}$ [1/s] & 55.325 & 54.899 \\ \hline
$\Omega_{21}^{90}$ [1/s]& 0.00676 & 0.00620 \\ \hline
$\Omega_{21}^{30}$ [1/s]& 937.1 & 1082.37 \\ \hline
$\Omega_{12}^{30}$ [1/s]& 0.037 & 0.046 \\ \hline
\end{tabular}
\end{table}

\section{Invariance of entropy production and efficiency}\label{sec:entropy}

An important question for any coarse-graining method concerns its effect on entropy production. In general, a coarse-grained description without imposed time-scale separation or detailed balance for the eliminated variables often underestimates the entropy production of the system \cite{raha07,nico11c,espo12,cris12,bo14}.
In this section, we show that for the type of models considered here, our coarse-graining method conserves the entropy production even if there is no time-scale separation between the eliminated and remaining degree of freedom.

Since transitions can be uniquely attributed to motor or probe particle, the total entropy production of the system \cite{seif12} can be split in two parts, analogous to bipartite or partially masked systems \cite{hart14,shir14}, 
\begin{align}
 &\stot=\sum_i\int_{-\infty}^{\infty} \frac{\gamma{j_i^x}^2(y)}{p_i(y)} \mathrm{d}y \nonumber \\
     & +\sum_{i,j,\alpha}\int\limits_{-\infty}^{\infty} p_i(y)w_{ij}^{\alpha}(y)\ln\frac{p_i(y)w_{ij}^{\alpha}(y)}{p_j(y+d_{ij}^{\alpha})w_{ji}^{\alpha}(y+d^{\alpha}_{ij})} \mathrm{d}y \nonumber \\
 &\equiv \stot^{\text{p}} + \stot^{\text{m}} \label{stot}
\end{align}
where $\jix=((\partial_y V(y)-\fex)p_i(y)+\partial_y p_i(y))/\gamma$ is the current due to the motion of only the bead for fixed $i$. Obviously, both $\stot^{\text{p}}$ and $\stot^{\text{m}}$ are non-negative.

The total entropy production (\ref{stot}) can be calculated using the LDB condition (\ref{eq:ldbmulti}) as
\begin{align}
 \stot =& \sum_i\int_{-\infty}^{\infty} (\partial_y V(y)-\fex) \jix \mathrm{d}y \nonumber \\ &+\sum_{i,j,\alpha}\int_{-\infty}^{\infty} p_i(y)w_{ij}^{\alpha}(y)\left(\Delta\mu_{ij}^{\alpha}-F_j+F_i\right. \nonumber \\
\qquad\qquad &\left.-V(y+d^{\alpha}_{ij})+V(y)\right) \mathrm{d}y\nonumber \\
 =&\sum_{i<j,\alpha}\Delta\mu^{\alpha}_{ij}j^{\alpha}_{ij}-\fex v \geq 0. \label{eq:avstot}
\end{align}
Using partial integration, it can be easily seen that the parts involving $V(y)$ cancel, i.e., the energy of the linker is constant on average. The total entropy production is then given by the chemical free energy consumption that is not transformed into mechanical power.

For the coarse-grained description, the total entropy production contains only contributions from the effective jump process,
\begin{align}
 \stot^{\text{cg}}=\sum_{i,j,\alpha} P_i\Omega^{\alpha}_{ij} \ln \frac{P_i\Omega^{\alpha}_{ij}}{P_j\Omega^{\alpha}_{ji}}%
=\sum_{i,j,\alpha}P_i\Omega^{\alpha}_{ij}\ln\frac{\Omega^{\alpha}_{ij}}{\Omega^{\alpha}_{ji}}. \label{eq:stotcg}
\end{align}
Using the LDB condition for the coarse-grained rates (\ref{eq:CGldbmulti}) and the condition on the operational current (\ref{occ}) yields
\begin{align}
&\stot^{\text{cg}}=\sum_{i<j,\alpha}\Delta \mu^{\alpha}_{ij}j^{\alpha}_{ij}-\fex v 
\label{eq:cgstot}
\end{align}
which is precisely (\ref{eq:avstot}). For these models for which the state space of the eliminated degree of freedom does not contain entropy producing internal cycles, the average total entropy production in the NESS remains invariant under our coarse-graining procedure. 

It is also instructive to apply the entropy-splitting scheme introduced in \cite{espo12} to our coarse-graining procedure. In \cite{espo12}, it was shown that the total entropy production can be written as a sum of the coarse-grained entropy production (\ref{eq:stotcg}) plus a contribution of the microstates corresponding to a mesostate (which are eliminated during coarse-graining) plus a contribution due to the fact that jumps between mesostates can occur involving different microstates.
In our framework, the total entropy production is already recovered by the coarse-grained entropy production. The two additional contributions which correspond to the total entropy production of the probe particle and the average total entropy production of the motor minus the coarse-grained entropy production cancel each other.

We finally show that our coarse-graining procedure also preserves the energy transduction, or thermodynamic, efficiency $\eta_{T}$ defined as the ratio of the extractable power $\dot{W}_{\text{out}}$ and the rate of chemical energy input $\dot{\Delta\mu}$ \cite{parm99},
\begin{align}
 \eta_{T}\equiv \frac{\dot{W}_{\text{out}}}{\dot{\Delta\mu}}.
\end{align}
For the systems we have studied so far, as long as the external force is smaller than the stall force, the power output is given by $\dot{W}_{\text{out}}=\fex v$ and the power input by $\sum_{i<j,\alpha}\Delta \mu^{\alpha}_{ij}j^{\alpha}_{ij}$ which leads to the efficiency
\begin{align}
 \eta_T=\frac{\fex v}{\sum_{i<j,\alpha}\Delta\mu^{\alpha}_{ij} j^{\alpha}_{ij}}
\end{align}
which is the same in the coarse-grained description since $v$, $j^{\alpha}_{ij}$ and $\Delta \mu^{\alpha}_{ij}$ are conserved.

For motor models with tight coupling or multi-state models with a single motor cycle, the rate of chemical energy input equals the velocity $\dot{\Delta\mu}=v\Delta\mu/d $ and the efficiency reduces to $\eta_T=\fex/\Delta\mu$. In general, however, any idle cycles of the motor increases the rate of chemical input over the velocity and therefore reduces the efficiency.

\section{Stall force and rate anomaly}
\label{sec:stall}

\begin{figure}[t]
\centering
\includegraphics[width=0.94\linewidth]{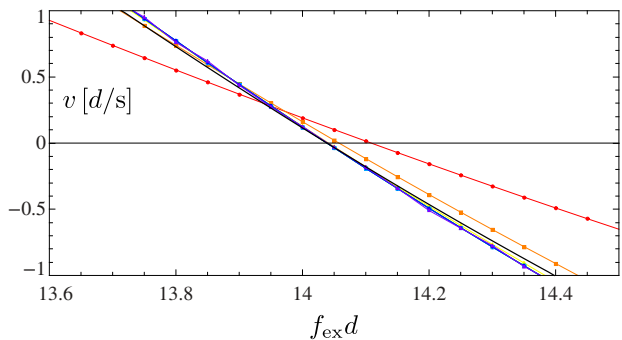}
\caption{(Color online) Average velocity of the kinesin model as a function of the external force. Depending on the size of the probe, stall conditions are reached for different values of the external force. Parameters as in Fig. \ref{OPOMmech}.}
\label{fig:stall}
\end{figure}

Coarse-graining multicyclic motor models as developed here reveals a remarkable feature concerning the stall force with significant implications on the interpretation of experimental data. For an example, consider the kinesin motor for the parameters chosen in Fig. \ref{OPOMmech}. Fig. \ref{fig:stall} shows that the stall force is a function of the size of the attached probe particle. Generally speaking, the stall force can indeed depend on the size of the probe since the network of the full system comprises more cycles than the coarse-grained or bare motor network, see Fig. \ref{fig:multi}. Varying the size of the probe, the relative weight of the cycles in the full system and hence their contribution to an operational current can change yielding a varying stall force. Thus, the experimentally obtained stall force corresponds to the stall conditions of the motor-probe complex but does not necessarily represent the stall conditions of the bare motor. If one is interested in the latter one should use very small probe particles since the limit of vanishing friction coefficient $\gamma$ is equivalent to applying the force directly to the motor. As discussed below, the stall force is independent of $\gamma$ for one-state or unicyclic multi-state motor models. Hence, an experimentally observed variation of the stall force with probe size can be used as proof that the motor is indeed multicyclic.

\begin{figure}[t]
\centering
\includegraphics[width=1.0\linewidth]{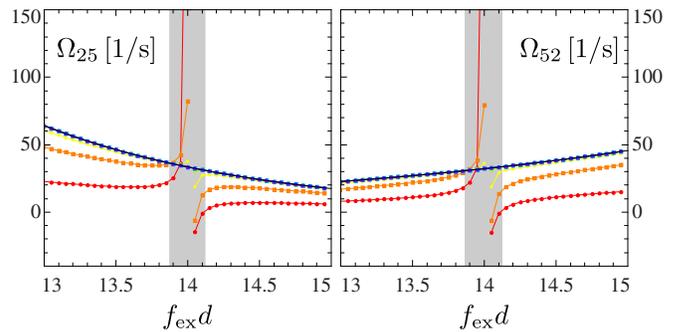}
\caption{(Color online) Detail of the coarse-grained mechanical transition rates of the kinesin model as shown in Fig. \ref{OPOMmech}. Near the stall force at $\fex d \simeq 14$ these rates exhibit a pole. In the gray shaded range, they should not be interpreted as physical transition rates.}
\label{OPOMmech_div}
\end{figure}

The varying stall force has also implications on the transition rates. In Fig. \ref{OPOMmech}, a close look around $\fex d=14$ shows that these data points are missing for the following reason. For all investigated models, we find that if, as a function of the external force, the sign change of an operational current depends on the friction coefficient $\gamma$, the coarse-grained rates corresponding to this transitions can become piecewise negative. This phenomenon occurs when the affinity of the affected transitions, $\ln\left[P_i\Omega^{\alpha}_{ij}/(P_j\Omega^{\alpha}_{ji})\right]$, has the opposite sign of $j_{ij}^{\alpha}$. An isolated sign change in the denominator of eqs. (\ref{OPmulti}, \ref{OMmulti}) leads to a pole in the
corresponding rate. Such an anomaly in $\Omega_{ij}^{\alpha}$ necessarily implies a
corresponding one in $\Omega_{ji}^{\alpha}$ since the ratio of the effective rates
obeys the local detailed balance condition which enforces the
same sign for both rates. In this range, the coarse-graining scheme introduced 
here fails to produce physically acceptable rates. In practice, one should 
discard the results at least when either a rate is negative or becomes 
larger than the rate for vanishing bead size. In Fig. \ref{OPOMmech_div}, where we zoom into the range around the stall force, this range
is shaded gray. Taken at face value, this phenomenon looks like a short-coming of our approach.
It is the price to pay for requiring over the full parameter range both 
the local detailed balance condition and the correct net currents from 
any one motor state to any other. While the negative rates do not allow for a sensible physical interpretation, they can nevertheless be used to calculate average quantities and yield, e.g., the correct entropy production as shown in section \ref{sec:entropy}.

This stall force anomaly with a corresponding range of negative rates occurs neither for any one-state motor model nor for unicyclic motors around stall conditions since only one motor cycle contributes to all cycles of the full system causing the zero of $j_{ij}^{\alpha}$ and $\ln\left[P_i\Omega^{\alpha}_{ij}/(P_j\Omega^{\alpha}_{ji})\right]$ to occur for the same $\fex$. We also found several multicyclic motors that do not lead to negative rates, e.g., the kinesin model if one assumes the chemical rates to be independent of $y$. A derivation of the precise 
conditions under which for multi-cyclic motor models a pair of effective rates 
diverges or becomes negative must be left to future work.  
We stress, however, that in all examples shown in this 
study, this anomaly occurs only in the narrow range shown in Fig. \ref{OPOMmech_div}. 
From a practical point of view, it may therefore not be as relevant as it is 
intriguing from a theoretical perspective.

\section{Conclusion}
\label{sec:conc}

Most experiments on molecular motors comprise some kind of probe particle. Therefore, any theoretical modelling with parameters estimated from experimental data will explicitly or implicitly contain characteristics of the probe particle. 

In this paper, we have introduced a systematic coarse-graining method that allows to reduce motor-bead models to effective one-particle motor models. This coarse-graining procedure provides a compromise between a one-particle description that is simple to handle and a detailed model comprising the dynamics of the full system. It yields an effective one-particle model maintaining the true motor network, where the influence of the probe is naturally incorporated without any additional assumptions since the simplification of the description takes place a posteriori. Any external force acting on the probe is then acting on the effective motor directly. The coarse-grained rates obey a LDB condition and yield the correct net currents. Fixing the marginal distribution and the average currents, there is still freedom on how to choose the rates. Only with the LDB condition the effective rates are determined uniquely. 

Applying the coarse-graining procedure to motor-bead models, we find that in general the coarse-grained rates do not show a single exponential dependence on the external force in contrast to what is often assumed for mechanical transition rates in one-particle models. Only in the often unrealistic limit of fast bead relaxation, the coarse-grained rates reduce to the corresponding one-particle rates.

In the absence of external forces, in general the coarse-grained rates are not proportional to the ATP concentration even if the motor rates obey mass action law kinetics. This feature originates from the drag effect of probe (due to friction) that is incorporated in the coarse-grained rates. For the same reason, the average velocity shows a sub-linear dependence on the ATP concentration even for a one-state motor model. Assuming an a priori one-state model with external force acting directly on the motor, one would have either to use a rather counterintuitive complex force-dependence of the transition rates or to introduce additional motor states in order to obtain a sublinearly growing velocity caused by the drag of the probe. In a one-particle description, the effect of large probe particles on the dwell time distributions could also be mistaken as signature of additional motor states thus leading to an overly complex motor network \cite{schi06,devi08}.

Considering the influence of the coarse-graining procedure on the stochastic thermodynamics of the system, we show that the total entropy production remains invariant under coarse-graining. This is due to the fact that, on the one hand, the state space of the eliminated degree of freedom contains no entropy producing cycles. On the other hand the design of the coarse-graining procedure is also important. It has to conserve the motor network as well as the net currents and provide transition rates fulfilling a LDB condition. Likewise, the thermodynamic efficiency remains invariant in our scheme. 

Our coarse-graining method conserves average quantities like the entropy production or operational currents although eliminating the dynamics of the probe particle strongly affects the cycle structure of the full system. In order to preserve also fluctuations of current observables in the long-time limit it was found that coarse-graining methods should conserve the cycle structure of the full system \cite{pugl10,alta12}.

From the experimental point of view, in order to obtain the simpler effective model, the underlying mesoscopic modelling need not to be known since all these quantities enter the coarse-grained description via the net currents and the marginal distributions which, in principle, can be extracted from the experimental data as we have demonstrated using a two step model for the $\text{F}_{1}$-ATPase.

The main advantage of the coarse-graining procedure introduced here is that once the rates have been obtained from experimentally accessible quantities, they automatically fulfill a LDB condition and provide the correct average currents, i.e., velocity, entropy production, hydrolysis rate etc..

For multicyclic motors, the coarse-graining procedure can yield rates that can have poles and become (piecewise) negative. If this scenario occurs, the coarse-grained rates lack a physical interpretation as transition probabilities in this range but they can still be used to calculate average quantities. For this class of motors the stall force typically depends on the size of the probe particle, i.e., the friction coefficient. Applying naively a one-particle model to such an experimental setup would not allow to determine the energy transduction mechanism of the motor correctly. For one-state motors, the coarse-grained rates are always positive. 

So far, we have discussed coarse-graining only under NESS conditions. In principle, the coarse-graining procedure as introduced in sections \ref{sec:CG-one} and \ref{sec:CG-mult} can also be applied to non-stationary states, e.g., if the nucleotide concentrations are not constant and $\Delta\mu$ decreases with time \cite{seif11,ge13}. Such a scenario would yield time-dependent $P_i$'s, net currents, LDB conditions and therefore also time-dependent coarse-grained rates.

Further generalizations might include other types of models representing the full system. While developed here for discrete motor models, the coarse-graining procedure should be also applicable to continuous motors moving in a tilted periodic potential where the potential minima will become the discrete states of the coarse-grained effective motor. The introduction of the index $\alpha$ in principle also accounts for more involved potentials or free energy surfaces that depend on both the motor and the probe state.

\begin{acknowledgments}
E.Z. thanks P. Pietzonka for valuable hints concerning numerical implementations. 
\end{acknowledgments}

\appendix*

\section{Limiting case: Large applied force}
\label{app:limit}

In the limit of large external forces, $\fex\rightarrow\infty$, the coarse-grained rates (\ref{OP}, \ref{OM}) can be expressed as
\begin{align}
 \op\approx&-v \exp[\Delta\mu-\fex d]/d \\
 \om\approx&-v/d.
\end{align}
While $\Delta\mu$ is independent of the external force, the average velocity is a function of the external force
\begin{align}
 v=\langle \partial_y V(y) -\fex\rangle/\gamma=\kappa\langle y \rangle/\gamma -\fex/\gamma.
\end{align}
It becomes negative for forces larger than the stall force $\Delta\mu/d$ which ensures that both $\op$ and $\om$ are positive. If there is no time-scale separation between the dynamics of motor and probe, $\av{y}$ grows linearly in $\fex$ for $\fex\rightarrow\infty$ with a smaller slope than $1/\kappa$. On the other hand with time scale separation, we have $\langle y\rangle =\fex/\kappa$. Note that within time scale separation, the average velocity has to be calculated using the average velocity of the motor, eq. (\ref{eq:vel}), since the ``average velocity'' of the probe $ \langle \partial_y V(y)-\fex\rangle/\gamma$ is zero as a result of the fast-bead limit of eq. (\ref{FPE}). Due to the linear dependence of $\av{y}$ on $\fex$, the average velocity, and therefore also $\om$, are then proportional to the external force whereas the exponential factor dominates for $\op$,
\begin{align}
 \op&\sim \fex\exp[\Delta\mu-\fex d] /(\gamma d)\\
 \om&\sim\fex/(\gamma d).
\end{align}

In the opposite limit of a large assisting force $\fex\rightarrow -\infty$, the coarse-grained rates (\ref{OP}, \ref{OM}) become
\begin{align}
 \op\approx&v/d \\
 \om\approx&v \exp[-\Delta\mu+\fex d]/d
\end{align}
As above, the average $y$ grows linearly and the velocity is proportional to $\fex$ if there is no time-scale separation which leads to
\begin{align}
 \op&\sim|\fex|/(\gamma d) \\
 \om&\sim|\fex|\exp[-\Delta\mu+\fex d]/(\gamma d).
\end{align}
This simple analysis clearly shows that the coarse-grained rates do not coincide with the often a priori assumed single exponential force-dependence of one-particle rates.
Within our numerical analysis, the asymptotic behaviour appears for $|\fex|\gtrsim 500/d$. The regime for large forces shown in Figs. \ref{OPOM} and \ref{OPOMct} is not the asymptotics yet. However, since $\av{y}$ is also linear in $\fex$ in this region yet with different slope, $v$ is still proportional to $\fex$.

\end{document}